\documentclass[]{aa}
\usepackage[varg]{txfonts}
\usepackage{siunitx}
\DeclareSIUnit\um{\micro\meter}
\DeclareSIUnit\Msun{M_{$\odot$}}

\usepackage{url}
\usepackage[tbtags,fleqn]{amsmath}
\usepackage{amsfonts}
\usepackage[x11names]{xcolor}
\usepackage{graphicx}
\usepackage{natbib}
\usepackage{booktabs}

\usepackage{dashbox,framed,color,ocg-p}
\newcommand{\e}{\mathrm e}

\newcommand{\mincir}{\raise
  -2.truept\hbox{\rlap{\hbox{$\sim$}}\raise5.truept \hbox{$<$}\ }}
\newcommand{\magcir}{\raise
  -2.truept\hbox{\rlap{\hbox{$\sim$}}\raise5.truept \hbox{$>$}\ }}


\newcommand{\teight}{$\tau_{850}$\ }
\newcommand{\likeli}{\mathcal{L}}

\begin{document}

\title{The HP2 survey}
\subtitle{III. The California Molecular Cloud: A Sleeping Giant Revisited \thanks{HP2 (\textit{Herschel-Planck}-2MASS) survey is a continuation of the series originally entitled "\textit{Herschel-Planck} dust opacity and column density maps" \citep{2014A&A...566A..45L, 2016A&A...587A.106Z}. Note: the flux, opacity and temperature maps presented in this paper are available in electronic form at the CDS via anonymous ftp to cdsarc.u-strasbg.fr (130.79.128.5)
or via http://cdsweb.u-strasbg.fr/cgi-bin/qcat?J/A+A/
 }}
\titlerunning{\textit{A Sleeping Giant Revisited}}
\author{Charles J.~Lada\inst{1}, John A. Lewis\inst{1}, Marco Lombardi\inst{1,2}, and Jo\~ao
  Alves\inst{3} }
\mail{marco.lombardi@unimi.it} \institute{%
  Harvard-Smithsonian Center for Astrophysics, Mail Stop 72, 60 Garden
  Street, Cambridge, MA 02138 \and University of Milan, Department of Physics, via Celoria 16, I-20133
  Milan, Italy  \and University
  of Vienna, T\"urkenschanzstrasse 17, 1180 Vienna, Austria 
  } \date{Received 22 May 2017; Accepted
  3 August 2017}

\abstract{
We present new high resolution and dynamic range dust column density and temperature maps of the California Molecular Cloud derived  from a combination of \textit{Planck}
and \textit{Herschel} dust-emission maps, and 2MASS NIR dust-extinction maps. 
We used these data to determine the ratio of the \SI{2.2}{\um} extinction coefficient to the
\SI{850}{\um} opacity and found the value 
to be close to that 
found in similar studies of the Orion B and Perseus clouds but higher than that 
characterizing the Orion A  cloud, indicating that variations in the fundamental optical properties of dust may exist between local clouds. 
We show that over a wide range of extinction, the column density probability distribution function (pdf) of the cloud can be well described by a simple power law (i.e., $\mathrm{PDF}_N \propto A_K^{\rm -n}$) with an index  ($n = 4.0 \pm 0.1$) that represents a steeper decline with $A_K$ than found ($n \approx 3$) in similar 
studies of the Orion and Perseus clouds. 
Using only the protostellar population of the cloud and our extinction maps we investigate the Schmidt relation, that is, the relation between the protostellar surface density, $\Sigma_*$, and extinction, $A_K$, within the cloud. 
 We show that $\Sigma_*$ is directly proportional to the ratio of the protostellar and cloud pdfs, i.e., $\mathrm{PDF}_*(A_K)$/$\mathrm{PDF}_N(A_K)$. We use the cumulative distribution of protostars to infer the functional forms for both $\Sigma_*$ and $\mathrm{PDF}_*$.
 We find that $\Sigma_*$ is best described by two power-law functions. At extinctions $A_K \lesssim \SI{2.5}{mag}$, $\Sigma_* \propto A_K^\beta$ with $\beta = 3.3$ while at higher extinctions $\beta = 2.5$, both values steeper than those ($\approx 2$) found in other local giant molecular clouds (GMCs). We find that $\mathrm{PDF}_*$ is a declining function of extinction also best described by two power-laws whose behavior mirrors that of $\Sigma_*$. 
 Our observations suggest that 
 variations both in the slope of the Schmidt relation and in the sizes of the protostellar populations between GMCs are largely driven by variations in the slope, $n$, of $\mathrm{PDF}_N(A_K)$. 
This confirms earlier studies suggesting that cloud structure plays a major role in setting the global star formation rates in GMCs} 

\maketitle

\section{Introduction}
\label{sec:introduction}

Located within the confines of the Perseus constellation, the California Molecular Cloud (CMC) rivals the Orion A molecular cloud as the most massive giant molecular cloud (GMC) within \SI{0.5}{kpc} of the Sun. However, despite its large mass, the CMC is a sleeping giant, characterized by a star formation rate (SFR) that is an order of magnitude lower than that of the Orion A cloud \citep{2009ApJ...703...52L}.  This fact along with its proximity make the CMC an ideal laboratory for investigating the physical process which regulates the global level of star formation in a cloud and sets the SFR.  Near-infrared extinction mapping provided the first complete maps of the dust column density structure of the CMC \citep{2009ApJ...703...52L, 2010A&A...512A..67L}.
Analysis of these observations and comparison with similar observations of the Orion~A cloud suggested a direct connection between the level of star formation activity in the cloud and its structural properties. In particular, it was suggested that the global SFR of the cloud was set by the amount of high extinction ($A_K \gtrsim \SI{1.0}{mag}$) material it contained  \citep{2009ApJ...703...52L}. 

In a subsequent study \citet{2009A&A...508L..35K} showed that it was generally the case that actively star forming clouds contained more high extinction material than did quiescent clouds.  Shortly thereafter \citet{2010ApJ...724..687L}  demonstrated that a linear correlation exists between the measured SFRs and  the gas masses at high extinction ($A_K \gtrsim \SI{0.8}{mag}$) for a nearly complete sample of molecular clouds within \SI{0.5}{kpc} of the sun. The connection between the SFR and high cloud surface densities was further strengthened by detailed studies of both local low mass and more distant high mass clouds \citep{2010ApJ...723.1019H,2014ApJ...782..114E, 2016ApJ...831...73V}.  There is also some evidence that this relation between SFR and dense gas extends smoothly to scales of entire galaxies \citep{2005ApJ...635L.173W, 2012ApJ...745..190L, 2013PASA...30...57J}. A further investigation of the Schmidt relation \citep{1959ApJ...129..243S} within individual local molecular clouds, including the CMC,
reinforced the idea that cloud structure was an important, if not the key, factor in determining the global SFR and level of star formation activity in a molecular cloud \citep{2013ApJ...778..133L}. 

In light of the above considerations it would be of great interest to examine in more detail the relationship between star formation and the highest extinction gas. However, this is not feasible using existing near-infrared extinction maps because such maps are severely limited in their ability to probe the highest extinction regions, due to the lack of background stars that can be detected in the presence of such high dust opacity.
More sensitive measurements of dust column density with higher dynamic range are required to investigate the nature of the star formation process at the high extinctions in which it occurs. With the dramatic improvements in the dynamic range of column density measurements provided by the {\textit Herschel} mission, we are now in a position to re-examine the detailed structure of the CMC and its relation to star formation. 

\begin{figure*}
 \centering
 \includegraphics[width=\hsize]{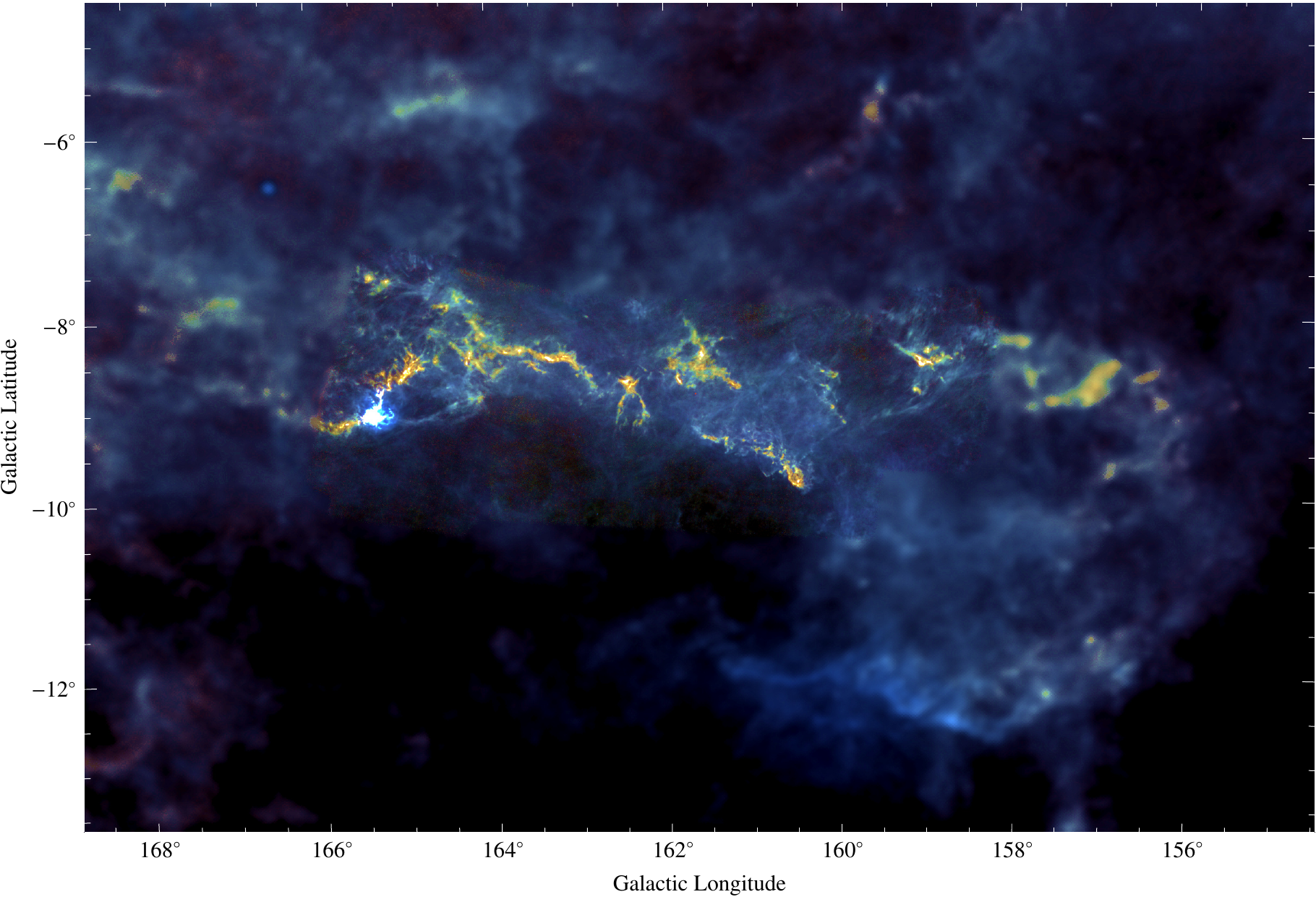}
  \caption{False color composite image showing the \SI{250}{\um} (blue), \SI{350}{\um} (green), and \SI{500}{\um} (red) SPIRE fluxes for the California Cloud region derived from \textit{Herschel} and \textit{Planck} observations. In the regions outside the \textit{Herschel} coverage a \textit{Planck/IRAS} dust model was used to predict the corresponding SPIRE fluxes. The spatial boundary between the \textit{Herschel} and the \textit{Planck} regions is readily identified due to the differing angular resolutions of the two data sets.}
  \label{fig:1}
\end{figure*}

A significant fraction of the CMC was surveyed by \textit{Herschel} \citep{2013ApJ...764..133H}. The surveyed region contained about half the cloud mass but importantly all regions of significant star formation. Here we analyze both \textit{Herschel} and \textit{Planck} observations of the CMC using the methodology of \cite{2014A&A...566A..45L}. This method combines the space observations with ground-based, near-infrared extinction maps to produce extinction-calibrated dust column density maps of a cloud. Such maps help mitigate the uncertainties in dust emission measurements of column densities and temperatures introduced by the well known degeneracy between these two physical quantities along a given line-of-sight. This is the third paper in a series applying this methodology to near-by clouds \citep{2014A&A...566A..45L, 2016A&A...587A.106Z}. The first two papers dealt with the more active star forming clouds Orion~A, Orion~B, and Perseus and demonstrated the power of the \textit{Herschel} observations for probing high extinction material and providing dust column density maps of GMCs with high dynamic range and angular resolution. The results of those papers provide a very useful data set for comparison with the present study of the CMC and will help elucidate the nature star formation in this relatively quiescent GMC.

The layout of this paper is as follows. Section~2 describes the data used for this study. Section~3 reviews the methodology used to produce the dust column density and temperature maps. Section~4 presents the results and in Section~5 we discuss the implications. A summary of our primary conclusions is found in Section~6.

\section{Data }
\label{sec:data}

The CMC was observed by the all-sky 
\textit{Planck} observatory and by the \textit{Herschel Space Observatory} as part of the ``Auriga-California'' program \citep{2013ApJ...764..133H}. 
The \textit{Herschel} data we used consisted of observations obtained in parallel mode simultaneously using the PACS and SPIRE \citep{2010A&A...518L...3G} instruments.
More details about the observational strategy can be found in \citep{2013ApJ...764..133H} and \citep{2010A&A...518L.102A}.  
We used the final data products of \textit{Planck} \citep{2014A&A...571A..11P}. 
Following the methodology of \citet{2014A&A...566A..45L} and \citet{2016A&A...587A.106Z}, the \textit{Herschel} data products were
pre-processed using the \textit{Herschel Interactive Processing
 Environment} \citep[HIPE;][]{2010ASPC..434..139O} version 15.0.1,
with the latest version of the calibration files.  The final maps were then produced using the HIPE level 2.5 of the SPIRE data, and the Unimaps data for PACS.

For the purposes of this study we use \textit{Herschel} observations made in the PACS \SI{160}{\um} band, and the SPIRE \SIlist{250;350;500}{\um} bands. 
However, prior to analyzing the data we multiplied each SPIRE band by an updated correction factor $C \equiv K_\mathrm{4e} / K_\mathrm{4p}$ \citep{2016A&A...587A.106Z} and 
then performed an absolute calibration of the \textit{Herschel} fluxes using \textit{Planck} maps. Next we 
convolved all \textit{Herschel} data to the resolution of the SPIRE \SI{500}{\um} band, 
i.e., $\mathit{FWHM}_{\SI{500}{\um}} = \SI{36}{arcsec}$.  

A false color image of the reduced \textit{Herschel} and \textit{Planck} SPIRE band fluxes in the CMC  region is shown in Figure~\ref{fig:1}.  The \textit{Herschel} observations cover the bulk of the molecular cloud in the central regions of the image. Here the three \textit{Herschel} SPIRE passbands have been convolved to the \SI{500}{\um} resolution of \SI{36}{arcsec}. The fluxes in the regions outside the \textit{Herschel} survey area are the predicted \textit{Planck} fluxes at the three  SPIRE passbands and are characterized by a much lower angular resolution of \SI{5}{arcmin}. We used the \textit{Planck/IRAS} dust model (i.e., T, $\beta$, $\tau_{850}$;  \cite{2014A&A...571A..11P}) to derive the expected fluxes in the SPIRE passbands. 
The boundary between the two regions is clearly apparent in the image because of the differing angular resolutions of the two surveys. We note here that although these two regions are clearly demarked by their differing angular resolutions, the colors in the high resolution region match the colors of the rest of the image very well indicating that the absolute calibration of the \textit{Herschel} observations is accurate. 

Finally, we used the NIR extinction maps of the CMC from 
\cite{2009ApJ...703...52L} that were derived from the 2MASS all-sky survey with the  \textsc{Nicest} technique \citep{2009A&A...493..735L}.

\section{Methodology}
\label{sec:methodology}

We derived the dust column density and temperature maps of the cloud following the methodology developed by \cite{2014A&A...566A..45L}.  This method was previously used to construct similar maps of the Orion~A and B clouds \citep{2014A&A...566A..45L} as well as the Perseus cloud \citep{2016A&A...587A.106Z}.  Here we briefly outline the general procedure. For more details the reader  can consult the Lombardi et al.\ paper.

\begin{figure*}[t]
 \centering
 \includegraphics[width=\hsize]{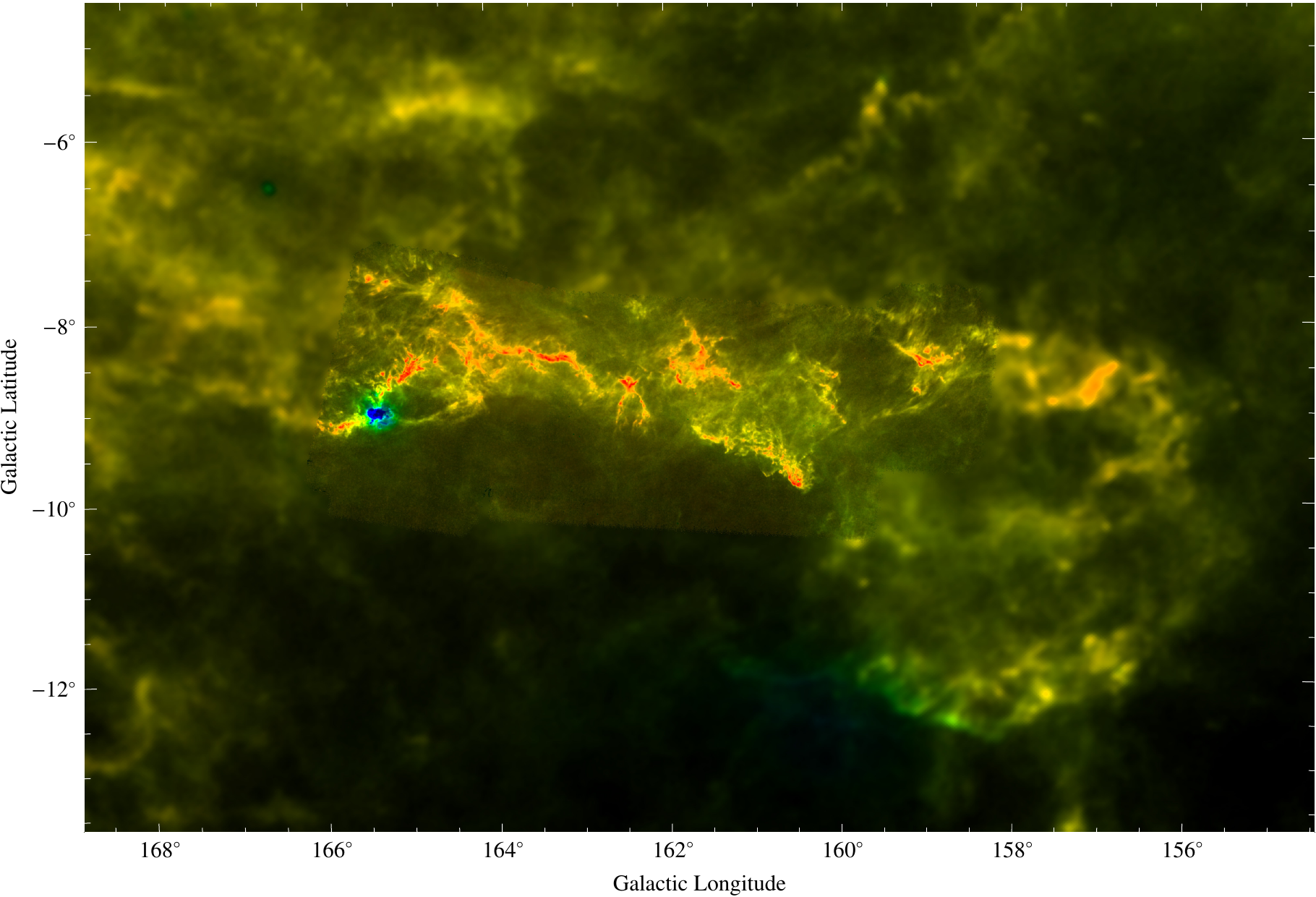}
  \caption{False color image showing the optical depth-dust temperature map for the California Molecular Cloud. The image shows optical depth as intensity and temperature as color with blue (red) corresponding to high (low) dust temperature. See text. }
  \label{fig:2}
\end{figure*}

\subsection{Physical model}
\label{sec:physical-model}

We consider the dust emission to be optically thin and describe the specific intensity at a frequency $\nu$ as a modified blackbody:
\begin{equation}
  \label{eq:1}
  I_\nu = B_\nu(T) \bigl[ 1 - \e^{-\tau_\nu} \bigr] \simeq B_\nu(T)
  \tau_\nu \; ,
\end{equation}
where $\tau_\nu$ is the optical-depth at the frequency $\nu$ and
$B_\nu(T)$ is the blackbody function at the temperature $T$:
\begin{equation}
  \label{eq:2}
  B_\nu(T) = \frac{2 h \nu^3}{c^2} \frac{1}{\e^{h \nu / k T} - 1} \; .
\end{equation}
Following standard practice, the frequency dependence
of the optical depth $\tau_\nu$ can be written as
\begin{equation}
  \label{eq:3}
  \tau_\nu = \tau_{\nu_0} \left( \frac{\nu}{\nu_0} \right)^{\beta_d} \; ,
\end{equation}
where  $\nu_0$ is an arbitrary reference
frequency. We set  $\nu_0 = \SI{353}{GHz}$, corresponding to $\lambda = \SI{850}{\um}$,
and we indicate the corresponding optical depth as $\tau_{850}$. 
This is also the standard adopted by the \textit{Planck}
collaboration. 

We note that when using this physical model we are implictly
assuming that temperature gradients are negligible along the line-of-sight. 
This is an approximation because non-negligible gradients in the dust temperature are clearly observed in the plane of the sky in many regions of molecular clouds. Therefore, it is likely that in such regions temperature gradients exist along the line-of-sight as well. 
Because of the sensitive dependence of  Eq.~\eqref{eq:1} on temperature, the temperatures derived from the observed fluxes using this equation will be  biased somewhat toward higher values. This will ultimately result in slight underestimates in the opacities and corresponding column densities. Therefore $T$ in Eq.~\eqref{eq:1} should be considered an \textit{effective dust temperature} for an observed dust column.

\subsection{SED fit}
\label{sec:sed-fit}

We can use our dust model to derive to derive the optical depth $\tau_{850}$, the effective dust temperature $T$, and the exponent $\beta_d$ in a given map pixel by fitting the modified blackbody of Eq.~\eqref{eq:1} to the fluxes measured by \textit{Herschel} for that pixel.  We use the reduced observations in the PACS \SIlist{100;160}{\um} bands, and the SPIRE \SIlist{250;350;500}{\um} bands to construct the spectral energy distribution (SED) at each map pixel.  We fit the modified blackbody dust model to the SED using a simple $\chi^2$ minimization that takes into account the calibration errors
(taken to be $15\%$ in all bands).
  Because of the degeneracies present in the $\chi^2$ minimization, we fixed $\beta_d$ and fit only for
  $\tau_{850}$ and $T$. For $\beta_d$ we adopted local values computed by the \textit{Planck}
  collaboration \citep{2014A&A...571A..11P} on 35 arcmin scales across the entire sky.

\section{Results}
\label{sec:results}

\subsection{Optical depth and temperature maps}

Figure 2 shows a combined optical depth and dust temperature map derived from the SED fits.
Here the image intensity is proportional to the opacity and color scales with temperature with blue representing hot ($T > \SI{25}{K}$) dust and red representing cold ($T < \SI{15}{K}$) dust. The coldest regions of the cloud (red) have dust temperatures  $\sim \SI{14}{K}$ while the hottest blueish region is characterized by temperatures between 18--\SI{33}{K}. These warmer temperatures are confined to the area around the the embedded cluster associated with the B star LkH$\alpha$~101 and the emission nebula NGC~1579. The dust in the bulk of the cloud is characterized by temperatures in the range 15--\SI{16}{K}. Optical depths range from roughly $10^{-4}$ to $10^{-3}$ with the highest values appearing closely associated with the coldest dust. Separate dust temperature and opacity maps for the region are shown in Figure~3.

\begin{figure*}[tp!]
  \centering
  \includegraphics[width=0.49\hsize]{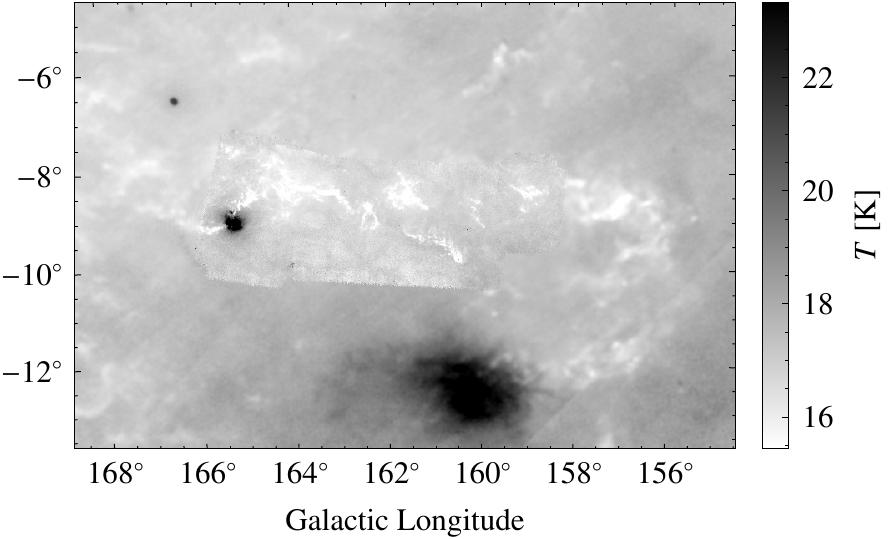}\hfill
  \includegraphics[width=0.49\hsize]{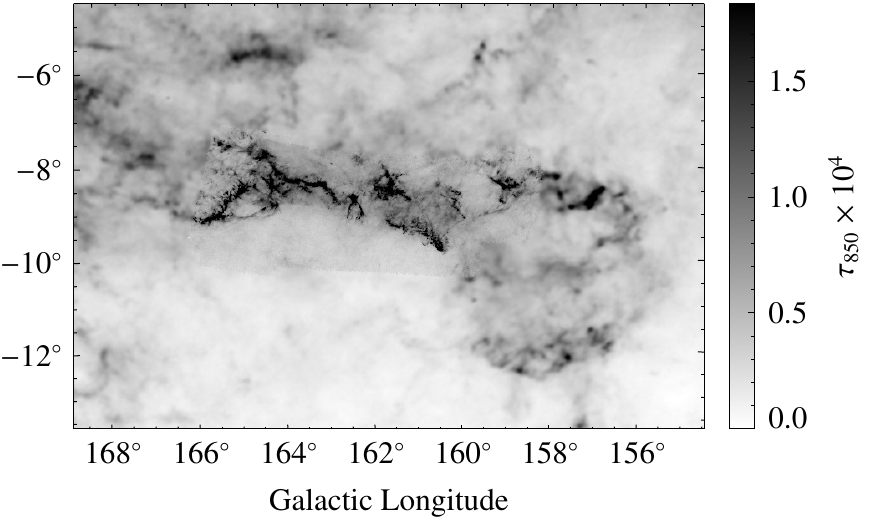}%
  \caption{Gray scale images showing the separate maps of dust temperature, $T_{D}$(eff),  (left) and optical depth, $\tau_{850}$, (right) for the California Molecular Cloud. The large dark region of high dust temperature at $l \approx \SI{160}{\degree}$ is the famous California Nebula, an HII region ionized by a runaway O star not associated with the California Molecular Cloud. The smaller, more compact hot region at $l \approx \SI{165.5}{\degree}$ is NGC~1579 and it is excited by the only embedded cluster in the cloud. Neither region appears noticeable on the optical depth map, illustrating the low opacities of the dust within the two HII regions.}
  \label{fig:3}
\end{figure*}
\subsection{Converting opacities to extinctions and column densities}
\label{sec:extinct-conv}

\begin{figure}[t]
 \centering
 \includegraphics[width=\hsize]{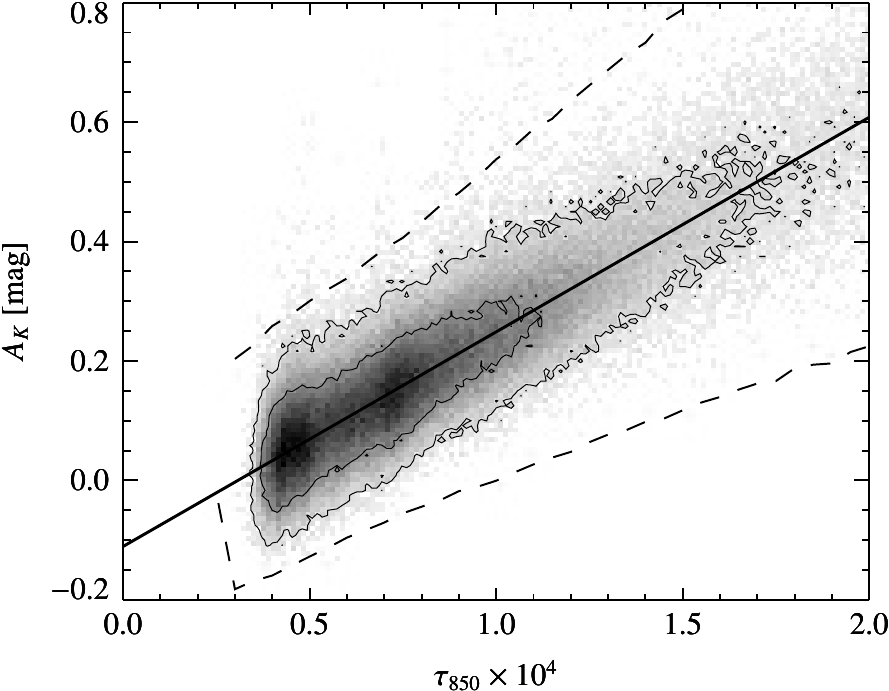}
  \caption{ Relationship between \textit{Herschel} $\tau_{850}$ and
    NIR extinction in the CMC.  The best linear fit, used for calibrating the opacities, 
    is shown as a black line. Dashed lines indicate the expected 3-$\sigma$ boundaries
    as calculated from direct error propagation in the extinction
    map. The contours enclose $68\%$ and $95\%$ of the points in the region, respectively. See text. }
  \label{fig:4}
\end{figure}

We can derive extinctions and ultimately useful column densities for the cloud from direct comparison of the $\tau_{850}$ and K-band (NICEST) extinction ($A_K$) maps. To make this comparison we considered only the \textit{Herschel} observations and first smoothed the \textit{Herschel} opacity map to the lower resolution (\SI{80}{arcsec}) of the extinction map. We empirically determined that between $0 \leq \tau_{850} \leq \num{2e-4}$ the relation between $\tau_{850}$ and $A_K$ is linear, 
that is:
\begin{equation}
  \label{eq:4}
  A_K(\mathrm{NIR}) = \gamma \tau_{850} + \delta \; .
\end{equation}
Note that since $\tau_\nu = \kappa_\nu \Sigma_\mathrm{dust}$, where, $\Sigma_\mathrm{dust}$ is the dust-column density and $\kappa_\mathrm{850}$ is the opacity coefficient (at \SI{850}{\um}), the slope $\gamma$ is proportional to the ratio of $\kappa_\mathrm{850}$ to $C_{2.2}$,
the extinction coefficient at \SI{2.2}{\um}. This follows from:
\begin{equation}
  \label{eq:6}
  A_K = -2.5 \log_{10} \left( \frac{I_\mathrm{obs}}{I_\mathrm{true}}
  \right) = (2.5 \log_{10} \mathrm{e}) \, C_{2.2}
    \Sigma_\mathrm{dust} \; .
\end{equation}
Thus, to first approximation, $\gamma \simeq 1.0857 C_{2.2} /\kappa_{850}$.  
The coefficient $\delta$ in Eq.\eqref{eq:4} absorbs calibration and systematic uncertainties that arise in both the \textit{Herschel} and extinction measurements. 
From fitting the linear relationship of Eq.~\eqref{eq:4} to the data the following values for the two parameters were derived:
\begin{equation}
  \label{eq:5}
  \begin{cases}
    \gamma =  3593 \pm \SI{75}{mag}  \\
    \delta = -0.110 \pm \SI{0.006}{mag}
  \end{cases}
\end{equation}
The data and fit are displayed in Figure \ref{fig:4}. The same figure also shows the predicted
3-$\sigma$ boundaries around the fit of Eq.~\eqref{eq:4}. These boundaries were
estimated from the statistical error on the extinction map alone (i.e., errors in the optical-depth were ignored). That the data points are predominately found to lie within 3-$\sigma$ boundaries indicates that
fit is accurate and confirms that the optical-depth map has a negligible relative 
error at the \SI{80}{arcsec} resolution of the extinction map.
The value of $\gamma$ we found for the CMC is higher than that obtained in a similar study of the Orion~A cloud (2640; \cite{2014A&A...566A..45L}), but similar to that derived for the Orion B cloud (3460; \cite{2014A&A...566A..45L}) and the Perseus cloud (3931; \cite{2016A&A...587A.106Z}).
As discussed in \cite{2014A&A...566A..45L} this suggests a possible variation in the optical properties of grains between local molecular clouds. The value of $|\delta|$ in the CMC is also higher than reported for the Orion and Perseus Clouds (i.e., $|\delta|= 0.001$--0.05). However, as pointed out by \cite{2010A&A...512A..67L} the NIR extinction measurements used here underestimate the true values by a systematic offset of \SI{-0.04}{mag} due the presence of an unrelated screen of foreground extinction that contaminated the 2MASS control fields used for this area of the sky. Once adjusted for that systematic error the value of $\delta$ is about \SI{-0.06}{mag}, comparable to those derived for the other regions.

For higher opacities the slope of the relation between $A_K$ and 
\teight flattens and the relation becomes non-linear. 
This departure from linearity and gradual flattening of the slope at higher extinctions is similar to that found previously for similar relations in Orion \citep{2014A&A...566A..45L} and Perseus \citep{2016A&A...587A.106Z}. 
It indicates that the infrared extinctions, $A_K$s,  are more often underestimating than overestimating the dust column densities in regions of high opacity. This shift from linearity is the likely result of a breakdown in the extinction technique at high column densities due to the relative paucity of background stars within the indivdual \SI{80}{arcsec} pixels of the 2MASS extinction map we used. 
This is confirmed by examining spatial maps of the differences, $A_K(\mathit{Herschel})-A_K(\rm{2MASS})$, which show the largest values confined to the regions of highest opacity, similar to what has been observed in the Orion \citep{2014A&A...566A..45L} and Perseus clouds \citep{2016A&A...587A.106Z}.

\begin{figure*}[t]
 \centering
 \includegraphics[width=\hsize]{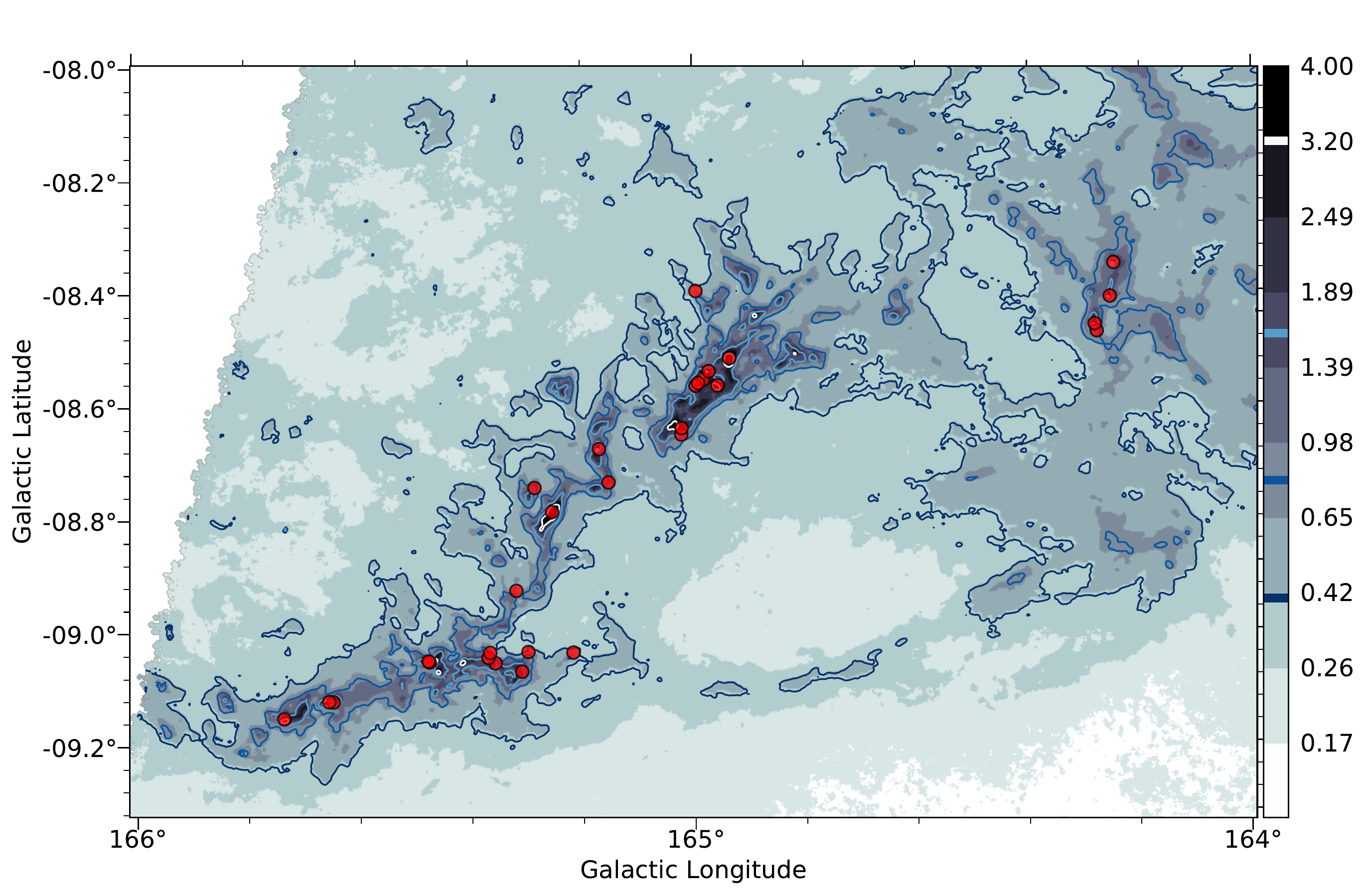}
  \caption{Zoomed-in, high resolution map of infrared extinction ($A_K$) derived from \textit{Herschel} observations for southeastern region (A) of the California Molecular Cloud. Fiducial extinction contours as well as the position of known protostellar objects (filled red circles) are drawn on top of the grey scale map. The contour values are 0.4, 0.8, 1.6 and 3.2 magnitudes and are indicated by thick colored lines superposed on the adjacent scale bar. The angular resolution is \SI{18}{arcsec}. See text.}
  \label{fig:regionA}
\end{figure*}

\begin{figure*}[t]
 \centering
 \includegraphics[width=0.9\hsize]{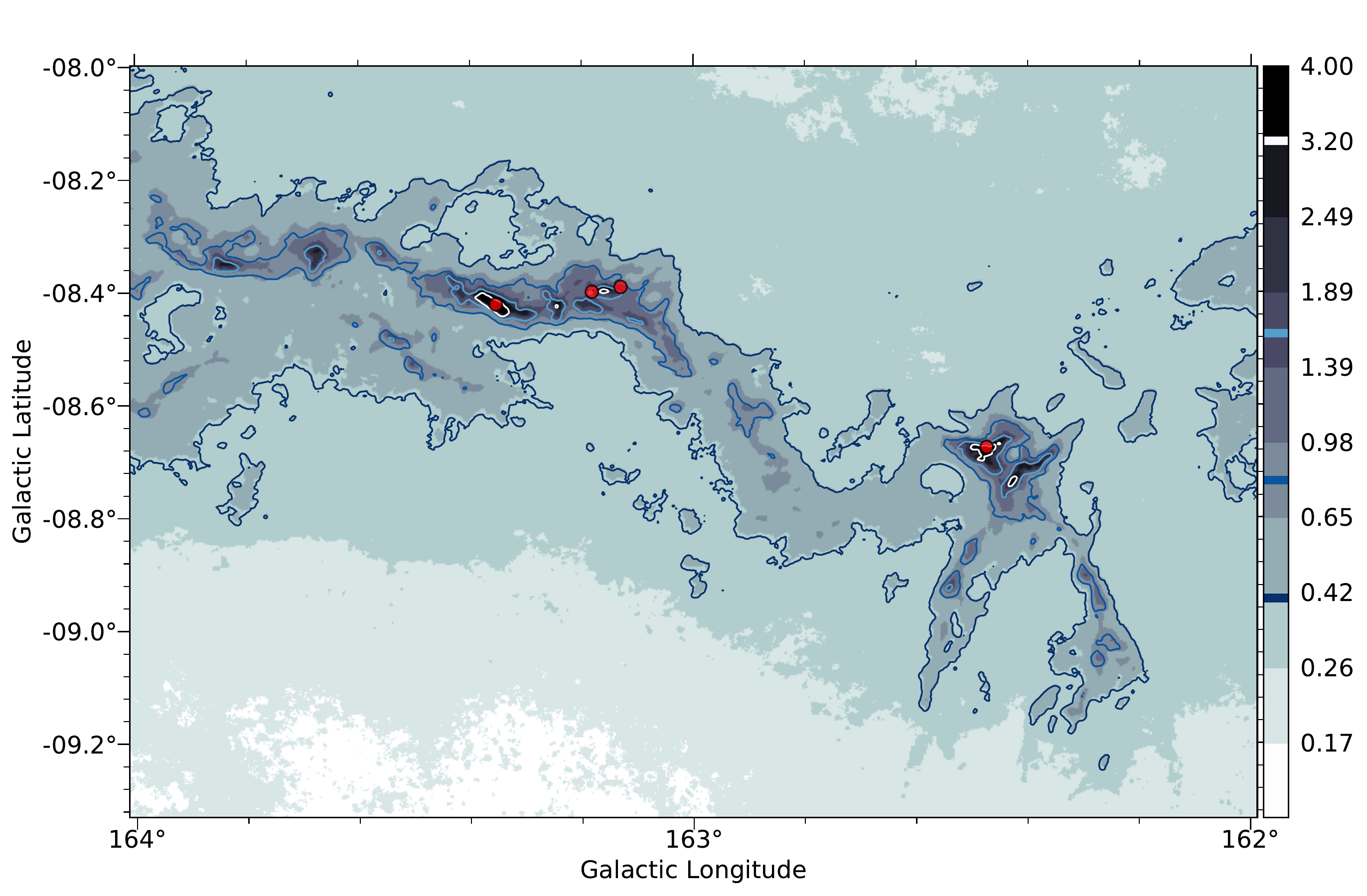}
  \caption{Zoomed-in, high resolution map of infrared extinction ($A_K$) derived from \textit{Herschel} observations for central region (B) of the California Molecular Cloud. Otherwise same as Figure~5. }
  \label{fig:regionB}
\end{figure*}

\begin{figure*}[t]
 \centering
 \includegraphics[width=\hsize]{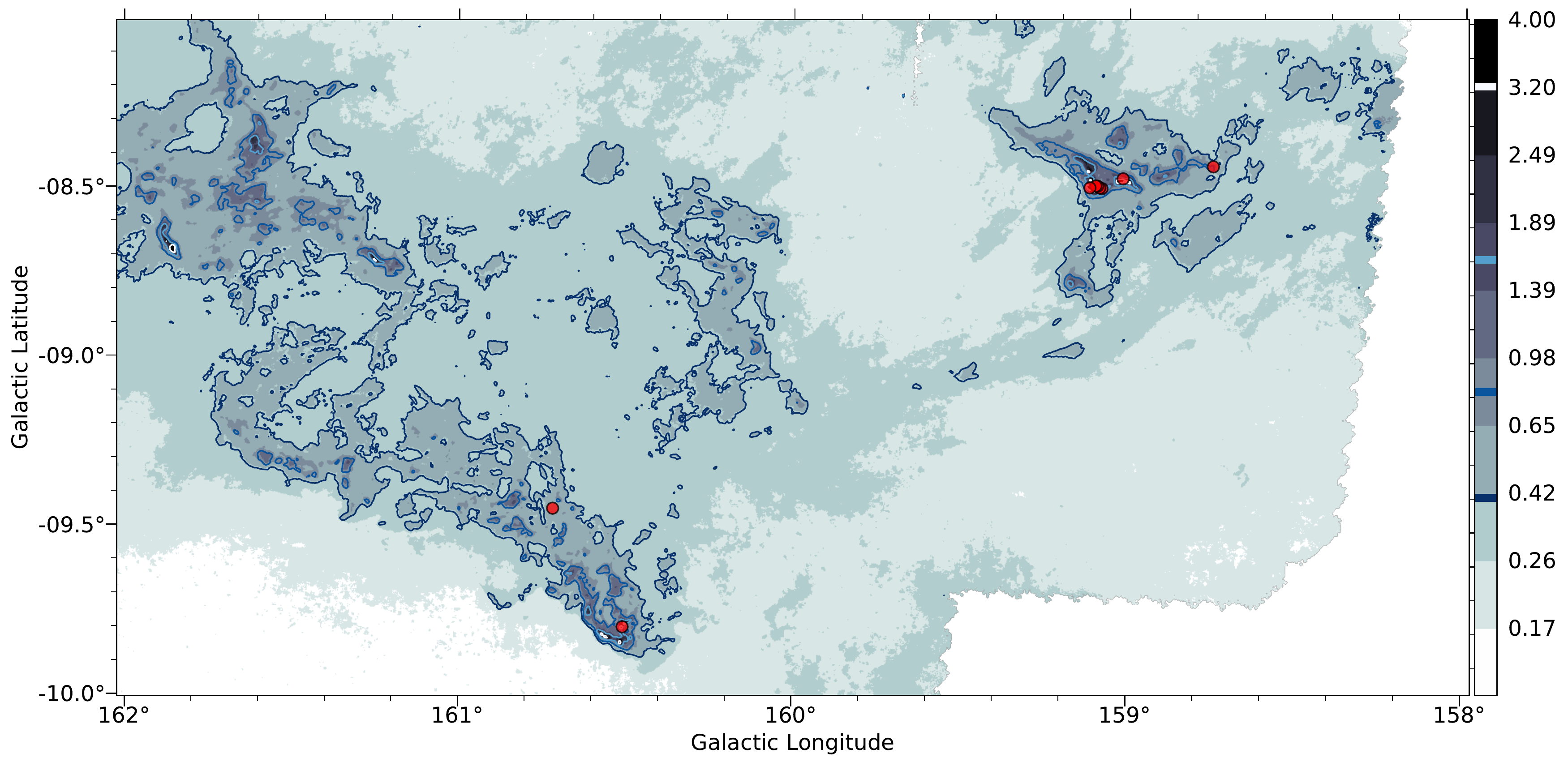}
  \caption{Zoomed-in map of infrared extinction ($A_K$) derived from \textit{Herschel} observations for northwestern region (C) of the California Molecular Cloud. Otherwise same as Figure~5. }
  \label{fig:regionC}
\end{figure*}

To construct a column density map of the region we convert the \textit{Herschel} submillimeter opacities to extinctions using only the coefficient $\gamma$, that is we set:
\begin{equation}
\label{eq:7}
A_K (\mathit{Herschel})= \gamma \tau_{850}  
\end{equation}   
In using equation \ref{eq:7} we have ignored $\delta$ because of its relatively small magnitude and the presence of a systematic bias present in the NIR extinction measurements as discussed above.  We applied equation \ref{eq:7} to the \textit{Herschel-Planck} opacity map to derive extinctions in each map pixel. We find a dynamic range in extinction of $0.04 \leq A_K \leq \SI{6.8}{mag}$ across our entire map (i.e., figure \ref{fig:2}).
In figures \ref{fig:regionA}, \ref{fig:regionB}, \ref{fig:regionC} we present zoomed-in maps of the extinction derived from Eq.~\eqref{eq:7} for three interesting sub regions (A, B, \& C) of  the CMC mapped by \textit{Herschel}.
These three maps were constructed at the higher spatial resolution (\SI{18}{arcsec}) of the SPIRE \SI{250}{\um} band following the methodology of \cite{2014A&A...566A..45L}.
Briefly, the SED of Eq.~(1) was fit  for $\tau_{250}$ in each SPIRE \SI{250}{\um} map pixel by using the flux in the \SI{250}{\um} band and fixing $T_\mathrm{D}$ to the value that was determined from the \SI{36}{arcsec} map (i.e., Figure~\ref{fig:3}) and fixing $\beta$ to the value derived by \textit{Planck} as before.  The value of $\tau_{850}$ was then derived from Eq.~(3). In the high resolution maps our ability to resolve small, high opacity features was improved with the maximum measured infrared extinction increasing from 6.8 to \SI{10.8}{mag}.

\subsection{Cloud Mass}

The extinctions derived from Eq.~\eqref{eq:7} can be converted to total (gas $+$ dust) mass surface densities from: 
$$
\Sigma_\mathrm{gas}  = \mu \alpha_K m_\mathrm{p} = \SI{183}{A_K} \ \ {\rm{M_\odot pc^{-2}}}
$$ 
Here, $\mu = 1.37$ is the mean molecular weight corrected for Helium,
$\alpha_K = \SI{1.67e22}{cm^{-2}.mag^{-1}}$ is the gas-to-dust ratio, $[N(\mathrm{HI}) + 2N(\mathrm{H}_2)]/A_K$
\citep{1978ApJ...224..132B, 1985ApJ...288..618R} and $m_\mathrm{p}$ is the mass of a proton. 
The total mass of the cloud is then obtained from integrating over the (\textit{Herschel} + \textit{Planck}) surface area of the cloud:
$$
M_\mathrm{tot} = \int \Sigma_\mathrm{gas} \, \mathrm{d}S
$$
To calculate the cloud mass we must first define the boundaries of the cloud. We consider the CMC to be entirely contained within a rectangular box on the sky given by $\SI{-13}{\degree} < b_\mathrm{II} < \SI{-6.5}{\degree}$ and $\SI{155}{\degree} < l_\mathrm{\rm II} < \SI{167}{\degree}$. We define the physical cloud boundary to be given by the $A_K > \SI{0.2}{mag}$ contour within the rectangular region.  This region includes both the inner area of the cloud surveyed by \textit{Herschel} as well as outer regions surveyed by \textit{Planck}. We applied Eq.~\eqref{eq:7} to both data sets to derive the dust column densities across this expanse of the CMC. We then find $M_\mathrm{tot} = \SI{1.1e5}{M_\odot}$. The region surveyed by \textit{Herschel} contains \SI{5.52e4}{M_\odot} or roughly half the cloud mass at $A_K > \SI{0.2}{mag}$. 

\section{Discussion}
 


\subsection{Protostellar Population and the Star Formation Rate}

Knowledge of the protostellar population of a cloud is instrumental in determining the relation between the star formation rate (SFR) and structure of the cloud \citep[e.g.,][]{2013A&A...559A..90L, 2013ApJ...778..133L}. 
Because of their relatively short lifetimes, the population of protostars  in a cloud can be considered the instantaneous yield of the star formation process in that cloud. Indeed, the number of protostars ($N_\mathrm{p*}$) is an excellent proxy for the instantaneous SFR since, to good approximation, $\mathrm{SFR} = m_\mathrm{p*} \tau_\mathrm{p*}^{-1} N_\mathrm{p*}$ where $\tau_\mathrm{p*}$ and $m_\mathrm{p*}$ are the typical lifetime and mass of a protostellar object.
Moreover, because protostellar objects are most likely to be within or very close to their original birth sites, their surface densities are the most appropriate for comparison with those of the gas or dust. Therefore, we consider here only protostars (i.e., Class~0, Class~I, and flat spectrum YSOs) in our determination of the cloud SFR.

\textit{IRAS} observations provided the first census of young stellar objects which displayed characteristics of potential protostars in the California cloud. Seventeen candidate protostars were identified from IRAS observations by \cite{2009ApJ...703...52L}.  Deeper \textit{Herschel} \citep{2013ApJ...764..133H} and \textit{Spitzer}  \citep{2014ApJ...786...37B} observations resulted in independent and more complete YSO catalogs of the cloud including greatly improved source classifications. These catalogs increased the known number of YSOs to 60 and 166 objects, respectively, including both protostars and more evolved (Class~II $+$ III) stars.  

Because the two catalogs were complied independently using slightly different criteria, there is some disagreement in source classifications between them.
For this study we merged these two catalogs and re-examined the source classifications following procedure outlined by \cite{Lewis:2016is} and summarized in the Appendix. 
Table~A.1 in the Appendix presents the merged catalog of YSOs we adopted. We identify 43 protostars in the CMC. 

The protostellar sources are largely confined to the portion of the cloud surveyed by \textit{Herschel}. The locations of 42 of these protostellar sources are shown on the extinction maps in figures \ref{fig:regionA}--\ref{fig:regionC}.  
These regions account for about 80\% of the cloud mass mapped by \textit{Herschel} and roughly 40\% of the total cloud mass in the \textit{Herschel} + \textit{Planck} maps.  Within the cloud boundary of $A_K > \SI{0.2}{mag}$ the protostellar surface densities are not uniform with $\Sigma_* = 0.2$, $0.03$ and \SI{0.02}{pc^{-2}} for regions A, B, and C, respectively. Region~A is considerably more active and efficient in forming stars than either B or C. Region~A contains $\sim$ 70\% of the protostars in the CMC while occupying only a small fraction (7\%) of the total (\textit{Herschel} + \textit{Planck}) cloud area and containing only a small fraction (10\%) of its total  mass. This region also contains NGC~1579 (LK~H$\alpha$101) the only substantial embedded cluster in this massive cloud. In all three regions the distributions of protostars closely follows the distribution of high extinction gas.

\subsection{Cloud Structure} 

\begin{figure}[t]
 \centering
 \includegraphics[width=\hsize]{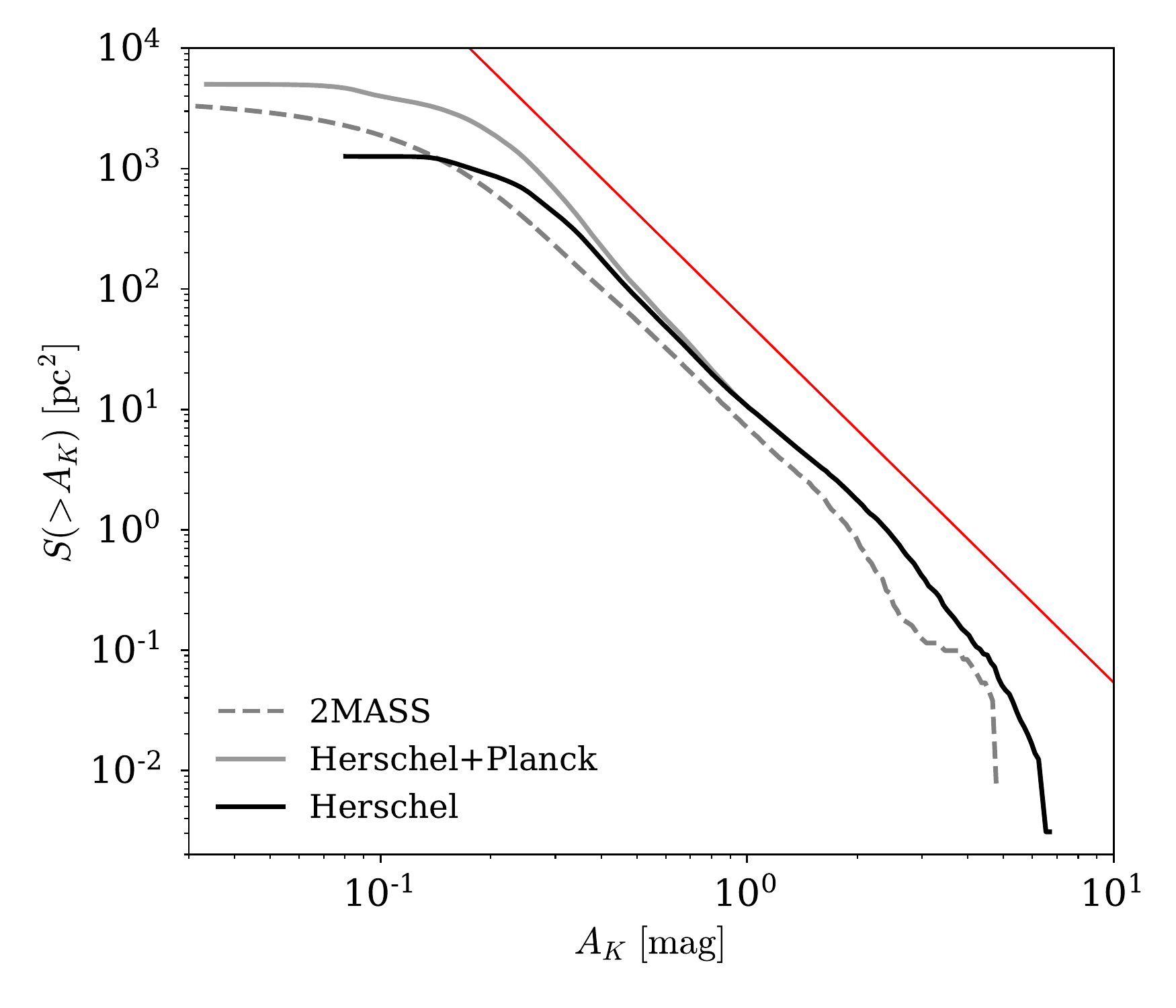}
  \caption{The surface area distribution functions, $S(>A_K)$, for the California cloud. The solid red line represents a power law relation with a slope of $-3$ for comparison. }
  \label{fig06a}
\end{figure}

The area distribution function is defined as 
$$
S(>A_K) \equiv \int_{A_K}^\infty \mathrm{d}S(A_K)
$$ 
where $\mathrm{d}S(A_K)$ is an element of cloud surface area at an extinction $A_K$. The area distribution function is a cumulative distribution that represents the total cloud area above a given threshold extinction as a function of $A_K$. The derivative of this function, $-S'(>A_K)$ is proportional to the column density PDF of the cloud and both functions are useful descriptors of cloud structure. Figure \ref{fig06a} and figure \ref{diffarea} show the functions $S(>A_K)$ and -$S'(>A_K)$ respectively. (We note that as plotted here the latter function corresponds to a linearly binned PDF and consequently its slope will differ by -1 from that of a logarithmically binned PDF, such as the ones studied by \cite{2015A&A...576L...1L}.) 
Although the \textit{Herschel} observations clearly extend to significantly higher extinctions than the 2MASS observations, in the range where the two data sets overlap, the two functions, $S(>A_K)$ and $-S'(>A_K)$ are in reasonable agreement with both data sets, confirming the result of our earlier study \citep{2013ApJ...778..133L} that both $S(>A_K)$ and $-S'(A_K$) fall off relatively steeply with $A_K$. For example, with $-S'(>A_K) \propto A_K^{-n}$ we find, from a formal fitting of the CMC data, that $n = 4.0 \pm 0.1$, for $A_K > \SI{0.35}{mag}$. This is relatively steep compared to the value ($\approx 3$) that describes the pdfs in Perseus and Orion.  Similarly for $S(>A_K) \propto A_K^{-q}$, where $q = n - 1 = 3.0$, the relation is relatively steep compared to that found ($\approx 2$) in the active star forming clouds Orion~A, Orion~B, and Perseus that we have also studied with \textit{Herschel}.


\subsection{The Schmidt Relation}

In previous studies of the CMC \cite{2013ApJ...778..133L} and \cite{2013ApJ...764..133H} showed that a Schmidt relation of the form $\Sigma_{*} \propto A_K^{\beta}$ existed for the cloud.  Here $\Sigma_{*}$, the surface density of protostellar objects and $A_K$ are proxies for $\Sigma_\mathrm{SFR}$ and $\Sigma_\mathrm{gas}$, respectively in the usual Schmidt relation \citep{1959ApJ...129..243S}.  However, using different methodologies and extinction data as well as slightly differing YSO catalogs they derived different values (2 and 4, respectively) for the index $\beta$.
In this paper we revisit the Schmidt relation for the CMC using a Bayesian (MCMC) methodology similar to \cite{2013ApJ...778..133L} coupled with our extinction calibrated, \textit{Herschel} \SI{36}{arcsec} resolution dust column density maps and our revised catalog of protostellar objects for the cloud.

We assume that the protostellar surface density behaves according to a thresholded Schmidt relation:
\begin{equation}
\label{eq:8}
\Sigma_{*}(A_K) = \kappa\, A_K^\beta\, \mathrm{H}(A_K - A_{K,0}),\ \
\end{equation}
where $\Sigma_*$ is the protostellar surface density, $A_K$ is the extinction in magnitudes in the pixel on which a source lies\footnote{It is important to note that here $A_K$  represents the total extinction along the line-of-sight to the protostar averaged over a \SI{36}{arcsec} or \SI{0.08}{pc} region and is not the pencil-beam extinction to the source itself.}, $\mathrm{H}(x)$ is a Heaviside function,  $A_{K,0}$ is the extinction threshold for star formation, and $\kappa$ is a normalization factor with units 
[\si{star.pc^{-2}.mag^{-\beta}}].
We do not model the diffusion of protostars from their birth location as was done in \cite{2013ApJ...778..133L} and \cite{2014A&A...566A..45L}. The diffusion, as previously measured, is at the sub-pixel level and is not degenerate with any parameter, so will not affect our determination of $\kappa$ or $\beta$. We take the likelihood derived in \cite{2013A&A...559A..90L} (their equation~7), $\likeli({x_n}|\theta)$, and estimate  $\hat\theta=(\kappa,\beta,A_{K,0})$ using the affine-invariant MCMC package \texttt{emcee} \citep{2013PASP..125..306F}, with the chains initialized to a Gaussian distribution around the result of a  Nelder-Mead ({\it amoeba})  maximization of the likelihood.

\begin{figure}[t]
 \centering
 \includegraphics[width=0.8\hsize]{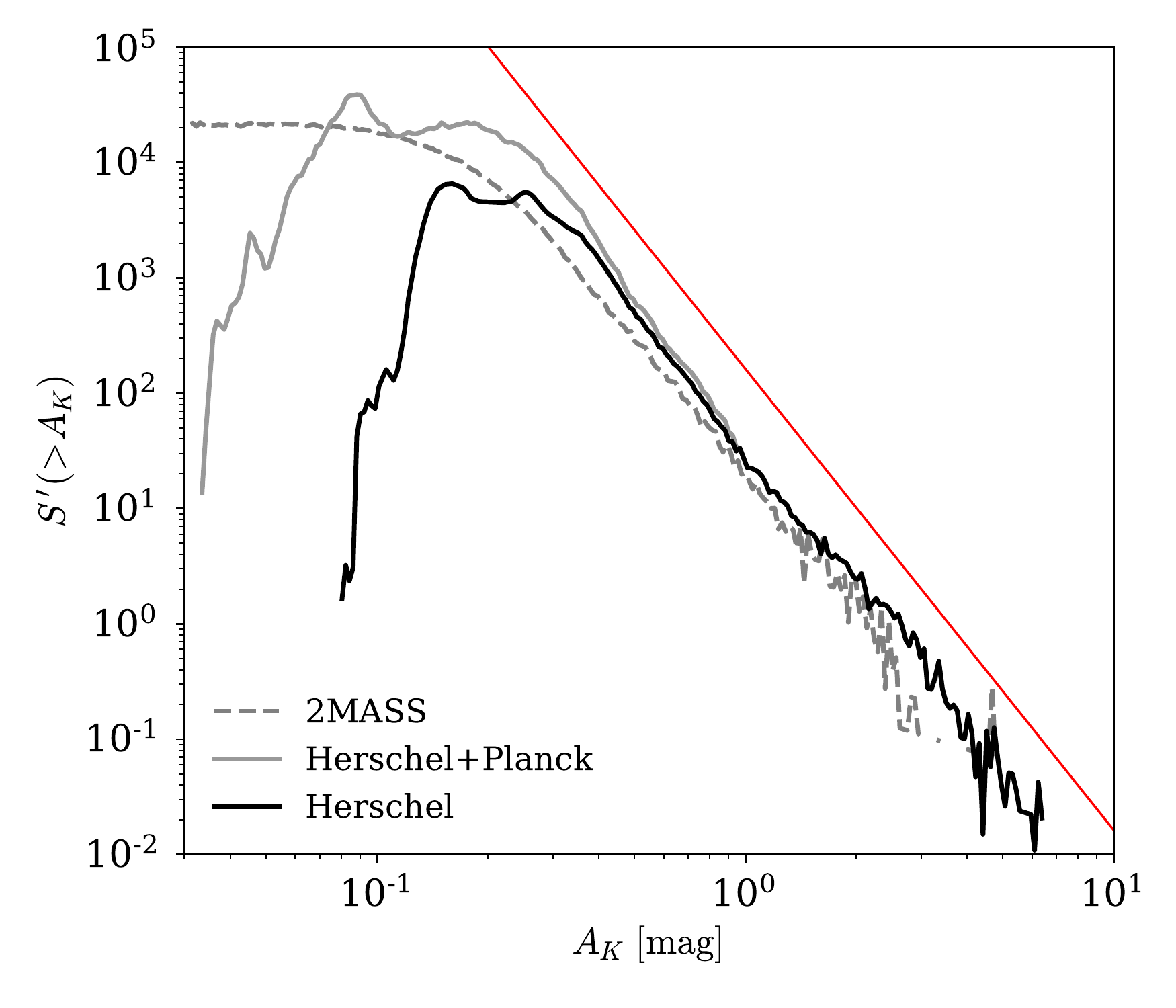}
  \caption{ The differential area function -$S'(>A_K)$ which is proportional to the probability density distribution for column densities in the cloud. In this plot a lognormal distribution would appear as a gaussian function whilst a power-law function would be a straight line. The red line shows the power-law $-S'(>A_K) \propto A_K^{-4.0}$.   See text. }
  \label{diffarea}
\end{figure}

\begin{figure}[t]
 \centering

 \includegraphics[width=\hsize]{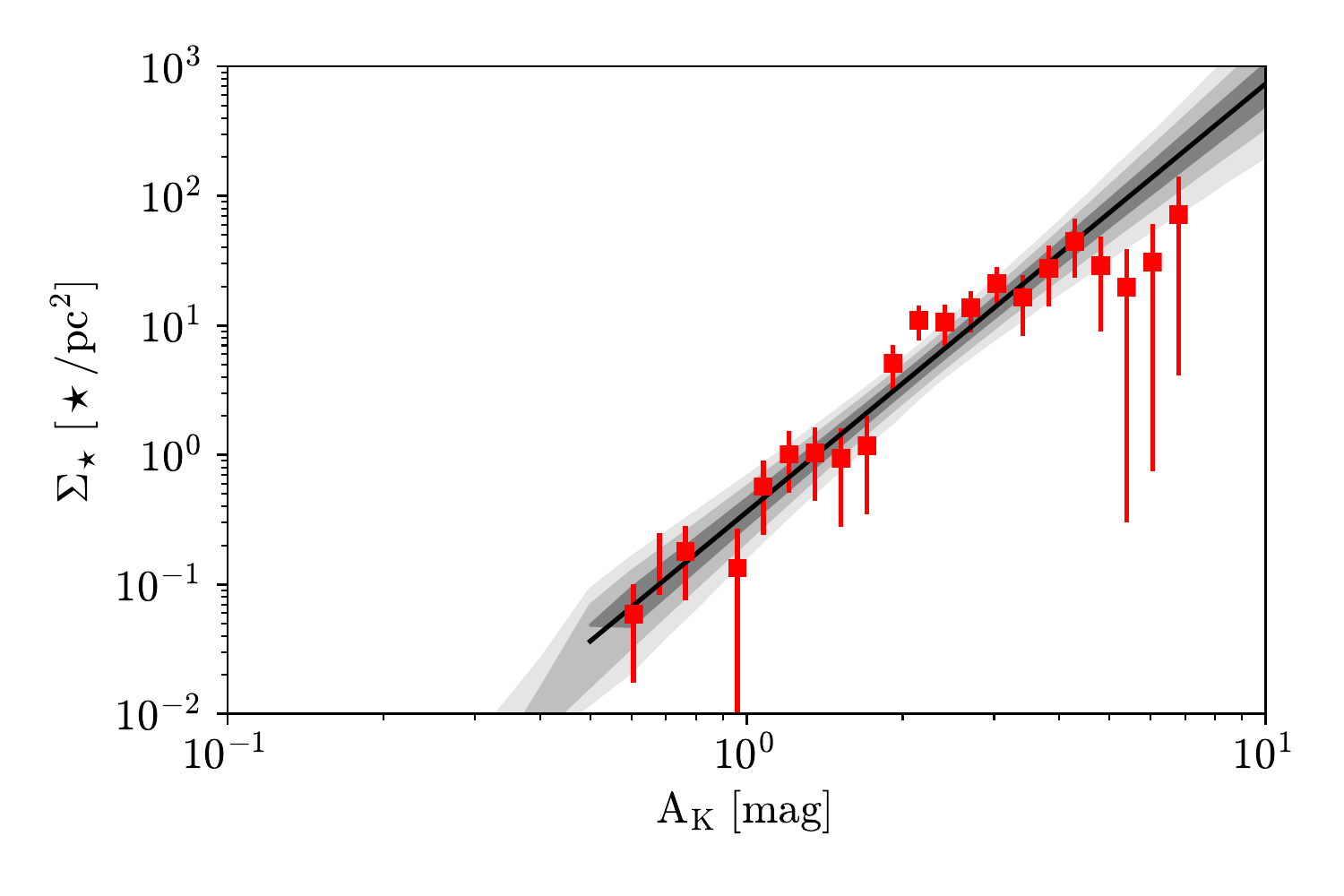}
   \caption{ The Schmidt relation for protostellar objects in the CMC. The data is shown as red squares with corresponding uncertainties. The black line represents the power-law relation, $\Sigma_{YSO} = \kappa A_K^\beta$ derived from the MCMC analysis with index $\beta$ $=$ 3.31. The gray shaded regions represent the 1, 2, and 3 $\sigma$ confidence levels for the plotted power-law relation. The data are fully sampled in logarithmic bins 0.1~dex wide equally spaced at intervals of 0.05~dex.  See text.}
  \label{slawCMC}
\end{figure}

From this analysis we determined the three credible intervals for the model in Eq.~\eqref{eq:8} to be $\beta = 3.31 \pm 0.23$, $\kappa = 0.36 \pm \SI{0.09}{stars.pc^{-2}.mag^{-3.31}}$, and $A_{K,0} = \SI{0.51}{mag}$. 
In figure \ref{slawCMC} we show the standard graphical representation of the power-law Schmidt relation we derived for the CMC plotted along with appropriately binned data for comparison. 
Visual inspection of the plot suggests that the fit and the model it is based on, i.e., Eq.~ \eqref{eq:8}, are more approximate than precise representations of the observations. In particular,
the data in the highest extinction bins clearly deviate from the $\beta = 3.31$ line. 
The departure of the highest extinction points from the fit suggests that a truncation may be present at high extinction. Such a truncation would likely result from the relative absence of high opacity material in the cloud as evidenced by the steep fall off of its PDF with $A_K$ (see Section 5.2 and Figure \ref{diffarea}).

%

The values of these posterior parameters differ significantly from those ($\beta \sim$ 2, $\kappa \sim 2$ and $A_{K,0} \sim \SI{0.6}{mag}$) derived by \cite{2013ApJ...778..133L} using similar methodology. This difference could arise from the different extinction maps and protostellar catalogs used in our two studies. The difference in the protostellar catalogs employed was slight and likely not responsible for the differing results. Nonetheless, we performed our analysis on various subsets of the original Harvey et al. catalog, including a version that used all sources. We found the resulting posterior parameters to be the same within the errors and thus not particularly sensitive to the different catalogs used. We next performed our analysis using the 2MASS NIR extinction map instead of our \textit{Herschel} map. In this case the analysis returned parameters that essentially reproduced the values derived by  \cite{2013ApJ...778..133L}. 
Closer inspection of the data showed that the extinctions associated with the individual protostars were almost always underestimated in the NIR map. This is not surprising since more than 80\% of the protostars in the CMC are found at $A_K > \SI{1.0}{mag}$ in the \textit{Herschel} extinction map. At these high opacities the NIR derived extinctions are considerably less reliable than our \textit{Herschel} extinctions and moreover are expected to underestimate the true opacities due to the small numbers of detectable background stars present in the individual pixels.  Our \textit{Herschel} derived value of $\beta$ is closer to, but slightly less than, that ($\sim 4$) derived by \cite{2013ApJ...764..133H} also using \textit{Herschel} observations.  Those authors employed a different methodology to estimate a value for $\beta$: First, \cite{2013ApJ...764..133H} produced surface density maps of dust ($A_K$) and YSOs, both smoothed to a scale of \SI{0.2}{\degree}. They then made a ratio of the two maps and constructed the corresponding plot of  $\Sigma_*$ vs $A_K$ from that data. 
\cite{2013ApJ...764..133H} quote only their value for $\beta$ derived from this plot and do not produce an estimate of its uncertainty so it is difficult to assess the significance of the difference between the two estimates. Because the value of $\beta$ derived using \textit{Herschel} dust column densities by two different methods is higher than the value we previously derived using 2MASS extinctions, we adopt our improved estimate for this parameter as being the more faithful measure of its true value.

\subsection{The Protostellar PDF}

\begin{table}
\caption{Schmidt relation and PDF power-law indicies for GMCs studied with \textit{Herschel}}
\label{slopes}
\centering
\begin{tabular}{c c c c c}
\hline\hline
GMC & $\beta$ & Reference & n & Reference\\
\hline
California & 3.31 $\pm$ 0.23 & 1 & 4.0 & 1\\
Orion A & 1.99 $\pm$ 0.05 & 2 & 2.9  & 4\\
Orion B & 2.16 $\pm$ 0.10 & 2 &  3.0 & 4\\
Perseus & 2.4 $\pm$ 0.6 & 3 & 2.7  & 4\\
\hline
\end{tabular}
\tablebib{
(1) This paper; (2) \citet{2014A&A...566A..45L}; (3) \citet{2016A&A...587A.106Z} (4) \citet{2015A&A...576L...1L}.
}
\end{table}

We can gain some physical insight into the nature of the Schmidt relation and its connection to star formation by writing the relation in the  following form:
 
\begin{equation}
\label{slaw_pdfs}
\Sigma_{*}(A_K)= \frac{\mathrm{d}N_{*}(A_K)}{\mathrm{d}S(A_K)} =  \Sigma_{*0} \times \frac{\mathrm{PDF}_{*}(A_K)}{\mathrm{PDF}_{N} (A_K)} 
\end{equation}
\noindent
 Here $\Sigma_{*0}$ is a constant equal to the global protostellar surface density of the cloud, that is, the ratio of $N_{*\mathrm{tot}}$, the total number of protostellar objects, to $S_{tot}$, the total cloud area. $\mathrm{PDF}_{*}(A_K)$ is the pdf of the protostellar population\footnote{i.e., PDF$_*(A_K) \equiv \frac{1}{N_{*tot}}\frac{dN_*(A_K)}{dA_K}$} and $\mathrm{PDF}_{N}(A_K)$ is the cloud column density pdf. 
 In this form we see that the Schmidt relation is proportional to the ratio of the protostellar and cloud pdfs. The column density pdf of a molecular cloud is a standard metric used to describe cloud structure. It has been shown to be well described by simple  power-law functions of extinction \citep[i.e., $\mathrm{PDF}_{N} \propto A_K^{-n}$;\ ][]{2015A&A...576L...1L}. The protostellar pdf is not a well known distribution and to our knowledge has not been studied in the literature. Because $N_*(A_K)$ is proportional to the total instantaneous SFR at any extinction, the protostellar pdf  is essentially the normalized measure of the SFR as a function of extinction. 
 Consider that we can rewrite equation \ref{slaw_pdfs} to yield:
\begin{equation}
\label{pspdf}
\mathrm{PDF}_*(A_K) =  \frac{\Sigma_*(A_K)}{\Sigma_{*0}} \times \mathrm{PDF}_N(A_K)
\end{equation}
If $\Sigma_*(A_K)$ and $\mathrm{PDF}_N(A_K)$  are both power-law functions of $A_K$, then $\mathrm{PDF}_*(A_K)$ must also be a power-law, i.e., $\mathrm{PDF}_*(A_K) \propto A_K^p$ where $p = \beta - n$. 

In Table~\ref{slopes} we list the values of $\beta$ and n derived from similar analysis of local clouds with \textit{Herschel} dust column densities. For all these clouds  $n > \beta$ indicating that $\mathrm{PDF}_*(A_K)$ is a declining function of $A_K$ and predicting that the number of protostars will steeply decline with extinction, despite the rapid rise of the Schmidt relation with $A_K$.
For the CMC we find $p = -0.69$ and for all four clouds $\langle \beta - n \rangle = -0.69 \pm 0.27$.
 
A protostellar pdf of the form $\mathrm{PDF}_* (A_K) \propto A_K^{-0.7}$ seems counterintuitive, in part because it clearly cannot describe the behavior of the actual $\mathrm{PDF}_*(A_K)$ at low extinctions. This is clear from figures \ref{fig:regionA}, \ref{fig:regionB}, \ref{fig:regionC}, which show an almost complete absence of protostars in regions of low extinction (i.e., $A_K \lesssim \SI{1.0}{mag}$).  
Moreover, it also predicts that the number of protostars will sharply decline with increasing extinction and this appears in conflict with the fact that the protostellar positions seem to be correlated with the highest extinction regions in the maps of figures \ref{fig:regionA}, \ref{fig:regionB}, \ref{fig:regionC}. 

 \begin{figure}[t]
 \centering
 \includegraphics[width=\hsize]{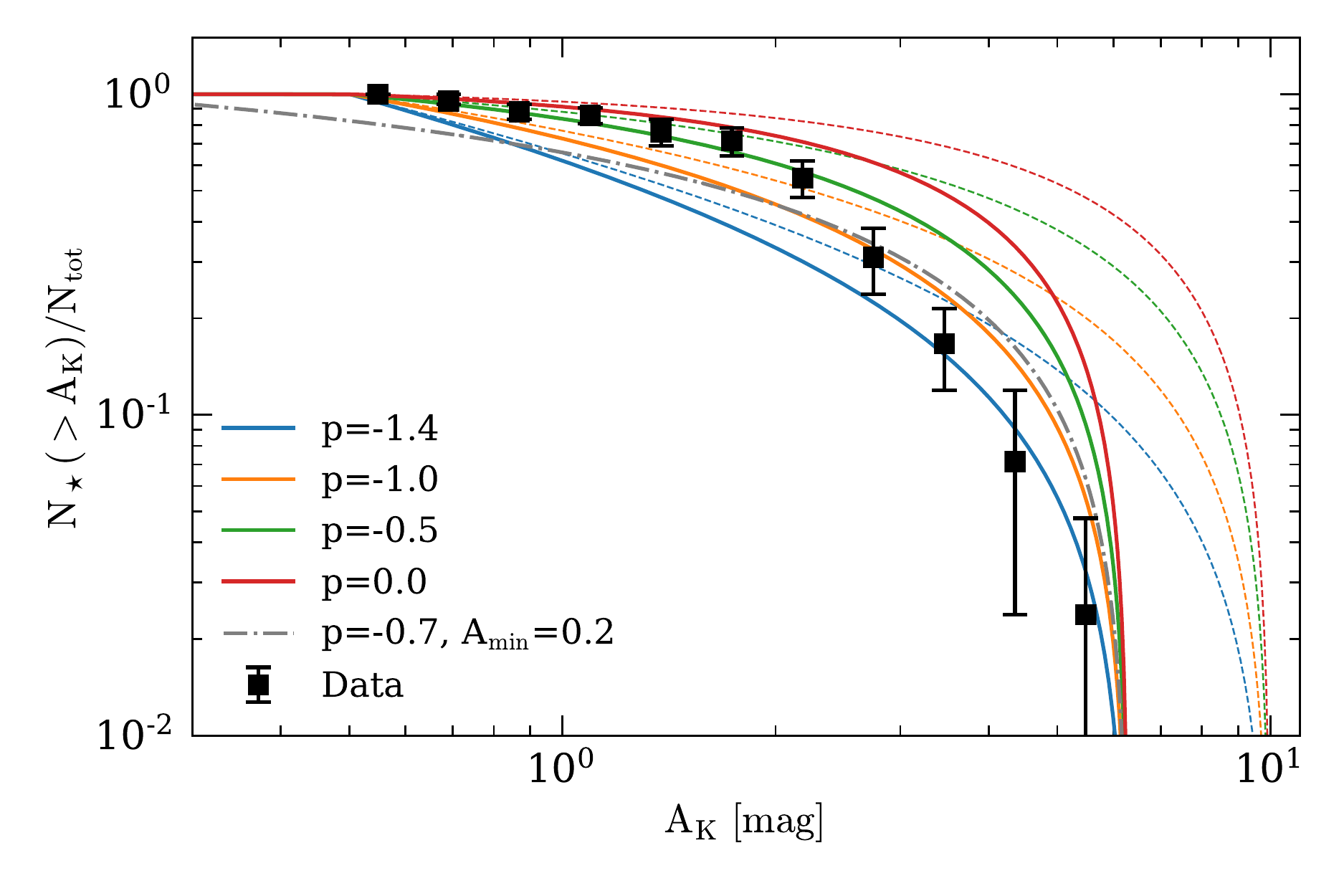}
  \caption{ The normalized cumulative distribution of protostellar objects with extinction in the CMC. The data are shown as filled squares with corresponding uncertainties. The continuous curves are the corresponding theoretical distributions expected for power-law protostellar pdfs i.e.,  $\mathrm{PDF}_*(A_K) \propto A_K^p$ and are truncated and normalized to 1.0 at $A_K = \SI{0.5}{mag}$. The solid curves correspond to $A_\mathrm{max} = \SI{6.3}{mag}$ and lighter dashed traces correspond to $A_\mathrm{max} = \SI{10.0}{mag}$. The dot-dash curve is normalized at \SI{0.1}{mag} to represent a cloud without an extinction threshold (see text).}
  \label{proto*pdf}
\end{figure}

 The relatively small numbers of protostars in the CMC coupled with the large dynamic range of extinction they sample make it difficult to directly determine a well sampled $\mathrm{PDF}_*(A_K)$ and test these predictions. 
 However, we can gain further insight into the nature of the protostellar pdf by considering the normalized cumulative distribution of protostars:
 \begin{equation}
 \label{cdp}
 N_*(>A_K)/N_{*tot} = \int_{A_K}^\infty \mathrm{PDF}_*(A_K) \mathrm{d}A_K  
 \end{equation}
This distribution can be considered the fractional yield of protostars as a result of the star formation process in the cloud. 
Figure \ref{proto*pdf} shows the normalized, cumulative distribution of protostars  as a function of extinction observed in the CMC. 
As predicted above, the observed data points (filled symbols) indicate that the number of protostars in the cloud in fact does sharply drop off with increasing extinction.  Similar results were found in the cumulative protostellar distributions in Orion~A, Orion~B, and Perseus. Moreover, the above analysis supports the hypothesis that, because $n > \beta$, the steep drop off of protostars at the highest extinctions is a direct result of the relative lack of such high extinction material in the cloud \citep[][and Figure \ref{diffarea}]{2013ApJ...778..133L}. 
 Figure \ref{proto*pdf} also confirms that rather than increasing, the number of protostars at low  extinctions becomes vanishingly small (i.e., $N_*(>A_K)/ N_\mathrm{tot} = 1.0$, for $A_K < \SI{0.5}{mag}$). 
Clearly the protostellar pdf cannot be described by a single power-law function that extends unabated to the lowest extinctions.

Using Eq.~\eqref{cdp} with $\mathrm{PDF}_*(A_K) \propto A_K^p$, the normalized cumulative distribution of protostars can be written as:
\begin{equation}
\label{intpdf}
{\rm N}(>A_K)/{\rm N_{tot}} =  (1 + p)C_p \int_{A_K}^{A_{max}} A_K^p dA_K.
\end{equation}
where ${C_p}$ is a normalization constant and $A_{max}$ is the extinction measured in the highest extinction map pixel containing a protostar. 
In Figure~\ref{proto*pdf} we also plotted a series of theoretical cumulative distributions given by Eq.~\ref{intpdf} for various values of the power-law index, $p$, of the protostellar pdf and two different values of the parameter $A_{max}$. 
To be consistent with the observations of the CMC we have normalized the functions to equal 1.0 at $A_K = \SI{0.5}{mag}$, effectively truncating the relations at that column density. For the solid set of curves we set $A_\mathrm{max} = \SI{6.3}{mag}$, the value derived from the data. In the dashed set of curves we set $A_\mathrm{max} = \SI{10.0}{mag}$. The solid curves provide the best match to all the data. 
 
Inspection of the figure confirms the notion that a single power-law $\mathrm{PDF}_*(A_K)$ is an inadequate description of the data.  However, at high extinction ($\gtrsim \SI{2.5}{mag}$) the  power-law with index $p = -1.4$ comes closest to matching the observations. In the range $0.5 \leq A_K \leq \SI{2.0}{mag}$, the data appear to best matched by the power-law with $p = -0.5$. From this simple comparison we can infer the general properties of the protostellar pdf in the CMC. At large extinctions $\mathrm{PDF}_*(A_K$) does appear to fall off as a power-law with index $\approx -1.4$. The function departs from this power-law at intermediate extinctions and appears to follow a shallower power-law with $p \approx -0.5$ between  0.5 and \SI{2.5}{mag} and then truncates or falls off very rapidly with decreasing extinction. This latter regime forms a broad peak in the pdf and accounts for roughly 80\% of the protostars in the cloud.  
 
 \begin{figure}[t]
 \centering
 \includegraphics[width=\hsize]{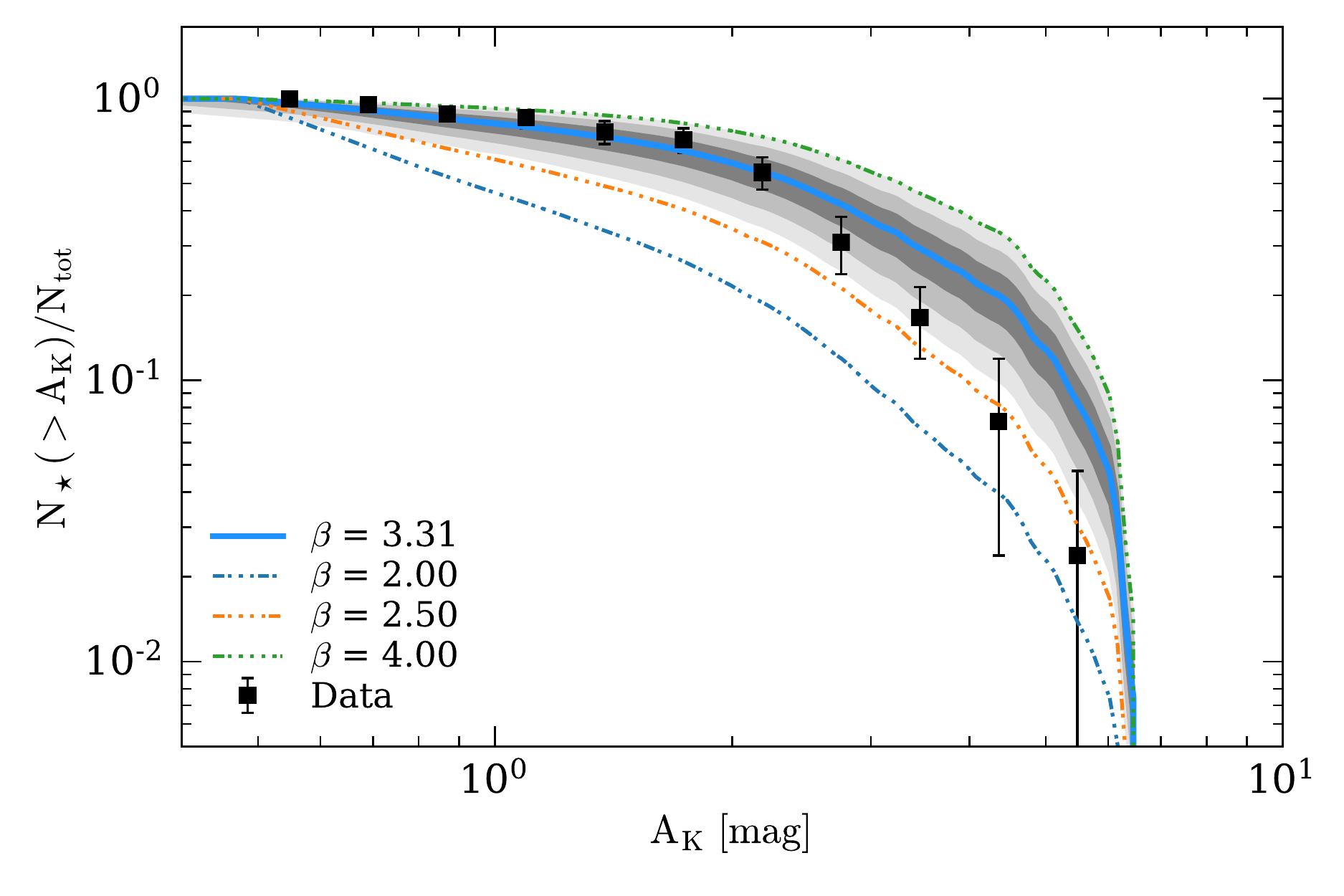}
  \caption{ The predicted distribution of protostars (solid line) calculated using the power-law relation, $\Sigma_{*} = 0.36 A_K^{3.31}$, derived from the MCMC analysis here along with the measured cloud PDF. The gray shaded regions represent the 1, 2, and 3 $\sigma$ confidence levels. The other lines show the predicted relations for different power-laws.}
  \label{ysofrac}
\end{figure}

 An instructive case is that of $p = 0$. Here there is equal probability of finding a protostar at any given extinction. In this case the indicies of the Schmidt relation and the cloud pdf would be the same, i.e., $n = \beta$. This is clearly not the case in the observations. That $\beta$ is found to be positive suggests that the star formation process operates more efficiently in gas at high extinctions but since $\beta < n$, not efficiently enough to prevent a decrease in the relative number of protostars formed (and the SFR) at the highest extinctions.
 
 The other set of curves plotted on figure \ref{proto*pdf} are cumulative distributions calculated from equation \ref{intpdf} with $A_\mathrm{max} = \SI{10.0}{mag}$.  These curves do not come close to matching the data and illustrate the sensitivity of the predicted
 distribution to the value of $A_\mathrm{max}$ and thus to $S'(>A_K)$, the un-normalized cloud column density pdf. For example, in the CMC, $A_\mathrm{max}$ is very close to the maximum observed extinction (\SI{6.8}{mag}) 
 in the \SI{36}{arcsec} pixels. However, the $A_\mathrm{max} = \SI{10}{mag}$ curves could correspond to a cloud with a $\mathrm{PDF}_N(A_K)$  that likely falls off more 
 slowly with extinction and as a result contains considerably more high extinction material than the CMC.  Comparison of these models with the data
again illustrates the importance of cloud structure in controlling the star formation process in the cloud.  

 Finally, we plotted a curve (dashed-dot trace) for a power-law with index $-0.7$ that is normalized at an extinction of \SI{0.2}{mag}. This illustrates the situation of a low extinction truncation that would correspond to the edge of a molecular cloud. This curve fails to match any of the observations. This reinforces the idea that there is a definite threshold for or truncation of the protostellar pdf at modest ($A_K \sim \SI{0.5}{mag}$) extinctions in the CMC. This truncation or threshold may be the result of an additional steepening (i.e., $\beta > 4$) of the Schmidt relation for $A_K < \SI{0.5}{mag}$ in the CMC.
  
 If the protostellar pdf cannot be described as a single power-law function then equation \ref{slaw_pdfs} requires that single power-law functions cannot describe the Schmidt relation and/or the cloud pdf. Inspection of figure \ref{diffarea} indicates that $\mathrm{PDF}_N(A_K)$ is very close to being a single power-law relation over a  large range (i.e., $A_K > \SI{0.2}{mag}$) in extinction. On the other hand, as discussed earlier and shown in figure \ref{slawCMC}, the Schmidt relation in the CMC is not particularly well fit by a single power-law function over a similar extinction range. Indeed, since the power-law indicies of the three functions are related as $\beta = p + n$, a change in power-law slope $p$ would be directly reflected by a change in $\beta$ for a constant $n$. At high extinctions we would expect $\beta = 4.0 - 1.4 = 2.6$ suggesting a flattening of the relation at high extinction as is seen in figure \ref{slawCMC}. At lower extinctions we would predict  $\beta = 4.0 - 0.5 = 3.5$, not far from the value of 3.3 that appears to fit the lower portion of the relation.

We can use the observed cumulative protostellar distribution function again to further constrain the Schmidt relation.
Figure \ref{ysofrac} plots the predicted distribution of protostellar objects obtained by numerically integrating:

\begin{equation}
\label{slawxds}
N_*(>A_K) = \int{\Sigma_*(A_K) \, \mathrm{d}S} = \int_{A_K}^\infty {\Sigma_*(A_K) \, |-S(>A_K)| \, \mathrm{d}A_K}
\end{equation} 
 The integration is carried out over the observed differential area function with $\Sigma_* \propto A_K^{\beta}$ for differing values of $\beta$. The uncertainties are shown for the case of $\beta = 3.31$ that we derived from the data assuming a model of a single power function. As observed in figures \ref{slawCMC} and \ref{proto*pdf} a single power-law does not appear match the observations. Allthough the low extinction data are well matched by the nominal $\beta \approx 3.3$ curve, the high extinction data in figure \ref{ysofrac} fall below the nominal $\beta = 3.3$ curve and are best matched by the $\beta \approx 2.5$ curve.  Both of these values match the prediction above, that $\beta = p + n$ with $n$ and $p$ independently determined from the data. These considerations imply that the Schmidt relation in the CMC is not described by a single power-law that rises with extinction but instead is a somewhat more complex function, more steeply rising in the outer regions of the cloud than in the inner high extinction regions. A qualitatively similar  behavior is observed in the cumulative distributions of protostars in the Orion~A, Orion~B and Perseus clouds and may be a general property of the internal Schmidt relation within a GMC. However, we note that the power-law indicies, $\beta$ and n, for the CMC are both greater than the corresponding indicies of the other three sources, which are very similar to each other. Yet the value of the index p is essentially the same for all four sources. This may imply a certain similarity of $\mathrm{PDF}_*(A_K)$ between the local GMCs.  A universal protostellar pdf would have interesting consequences for star formation theory. 
For example, equation \ref{slaw_pdfs}  would then suggest that cloud-to-cloud variations in the slope of the Schmidt relation are primarily driven by variations in the slope of the cloud column density pdf.



\section{Conclusions} 
\label{sec:conclusions}

 We have constructed high-resolution, high dynamic range dust column-density and
temperature maps of the California Molecular Cloud. The maps were derived by fitting \textit{Herschel} fluxes in each map pixel with a modified blackbody to derive the dust opacities and effective (i.e, line-of-sight weighted) temperatures. The opacities were calibrated at low extinction by 2MASS NIR extinction measurements to produce final maps of dust column density expressed as $A_K$. The column-density maps span a high dynamic range covering $\SI{0.04}{mag} < A_K < \SI{11}{mag}$, or $\SI{6.7D20}{cm^{-2}} < N < \SI{1.8D23}{cm^{-2}}$, a considerably larger range than measured in previous NIR extinction maps.  
This enables us to investigate cloud structure and star formation to much greater depths in the cloud than previously possible.

We used these data to determine the
  ratio of the \SI{2.2}{\um} extinction coefficient to the
  \SI{850}{\um} opacity and found the value ($\approx$ 3600) to be close to that (3500) found in a similar study of the Orion B cloud but higher than that ($\approx 2500$) characterizing the Orion A cloud, indicating that variations in the fundamental optical properties of dust may exist between local clouds.  
  
 We find that the column density pdf of the cloud can be well described over a large range of extinction ($0.35 \lesssim A_K \lesssim \SI{11}{mag}$)  by a simple power law (i.e., $\mathrm{PDF}_N(A_K) \propto A_K^{-n}$) with an index  ($n = 4.0 \pm 0.1$) that represents a steeper decline with $A_K$ than found ($n \approx 3$) in similar \textit{Herschel-Planck} studies of the Orion~A, Orion~B, and Perseus clouds. 
  
 We re-examined existing catalogs of YSOs in the cloud and produced a merged catalog with slightly revised classifications for the known YSOs. Using only the protostellar population of the cloud and our \textit{Herschel} extinction maps we investigated the Schmidt relation within the cloud. If we assumed that this relation is given by a simple power-law that is,  $\Sigma_* \propto \Sigma_\mathrm{gas}^\beta$, we found $\beta = 3.31 \pm 0.23$, a slope that is significantly steeper than that ($\approx 2$) found in other local GMCs. However modeling the cumulative distribution of protostars in the cloud indicated that $\Sigma_*$ is not a simple power-law function. Instead it is better described by two power-laws. At low extinction it is a rapidly rising power-law with an index of 3.3, while at higher extinctions it is characterized by a more slowly rising power-law with an index $\approx 2.5$. 
 
 We showed that $\Sigma_*$ is directly proportional to the ratio of the protostellar pdf, $\mathrm{PDF}_*(A_K)$, and $\mathrm{PDF}_N(A_K)$ and that if $\Sigma_*$ and $\mathrm{PDF}_N$ are simple power-law functions,  $\mathrm{PDF}_*(A_K)$ must be a simple power-law as well, that is, $\mathrm{PDF}_*(A_K) \propto A_K^p$ where $p= \beta - n$. 
 
 We used the cumulative distribution of protostars to constrain the functional form of  $\mathrm{PDF}_*(A_K)$. We found that it is not well described by a single power-law. At high extinctions $\mathrm{PDF}_*(A_K)$ is a declining power-law with $p \approx -1.4$. Between $0.5 \lesssim A_K \lesssim \SI{2.5}{mag}$ the function is characterized by a shallower power-law with $p \approx -0.5$. Below \SI{0.5}{mag} it appears to be truncated and must fall steeply with decreasing extinction. Its behavior closely mirrors that of $\beta$, exactly as expected if  $\mathrm{PDF}_N$ is well described by a single power-law function. Furthermore, our observations tentatively suggest that $\mathrm{PDF}_*(A_K)$ does not vary significantly between local clouds. If so, the variation in  $\beta$  between the CMC and other nearby GMCs is largely driven by differences  in the slope, $n$, of the column density pdfs of these clouds. However because the magnitude of the spectral index, $n$, of the cloud pdf is greater than that of the Schmidt relation, $\beta$, the size of the protostellar population and thus the global SFR in the cloud is largely controlled by the cloud's structure  (i.e., $\mathrm{PDF}_N(A_K)$).

Finally, our observations of the CMC have provided valuable insights into the star formation process by adding to the evidence that cloud structure is critical to setting the level of star formation in a cloud. However, we have not explained why the CMC has the structure it does. Although this was likely inherited from its formation, we not do not know whether the CMC will remain a sleeping giant or awaken and evolve its structure to resemble the Orion clouds with considerably more high extinction material and the corresponding increased levels of star formation.

\begin{acknowledgements}
 We thank Jan Forbrich for informative discussions.  Based on observations obtained with \textit{Planck}
  (\url{http://www.esa.int/Planck}), an ESA science mission with
  instruments and contributions directly funded by ESA Member States,
  NASA, and Canada.  J.~Alves acknowledges support from
  the Faculty of the European Space Astronomy Centre (ESAC) and J. Lewis acknowledges support from NSF GRFP grant DGE1144152.
\end{acknowledgements}


\bibliographystyle{aa}

\begin{appendix}

\section{The YSO Catalog for the California Molecular Cloud}

The two most extensive  catalogs of YSOs in the CMC are based on \textit{Spitzer, Herschel}  and \textit{Wise} data  \citep{2013ApJ...764..133H, 2014ApJ...786...37B}.
\citeauthor{2014ApJ...786...37B} selected {\textit Spitzer} YSOs using color-color and color-magnitude diagrams and then used \textit{Herschel, Spitzer} and \textit{WISE} observations to construct SEDs with sufficient wavelength coverage to provide suitable classifications of the objects. The $2-24~\mu{\rm m}$ SED slope was employed to obtain their YSO classifications, producing 58 protostellar objects including objects classified as flat-spectrum sources. \citeauthor{2013ApJ...764..133H} used SED slopes from $3.6-160~\mu{\rm m}$ to classify the \textit{Herschel} sources in the CMC, and used the bolometric temperature ($T_{\rm bol}$) over the same range to separate Class 0 ($T_{\rm bol}\le 50 {\rm ~K}$) and Class I ($50 {\rm ~K} \le T_{\rm bol} \le 70 {\rm ~K}$) sources. They found 49 sources with protostellar like SEDs
Because the two analyses are  independent, there is some disagreement between the catalogs in classifications of the same sources. By combining both data sets we create better, more complete, spectral energy distributions to use for source classification. We re-examined the source classifications following the methodology of \cite{Lewis:2016is}.
We first ensured proper matching by requiring that every object \citeauthor{2013ApJ...764..133H} matched to the \citeauthor{2014ApJ...786...37B} catalog \citep[see][Table 2]{2013ApJ...764..133H} is matched in our merged list, and we resolve conflicts by examining every match by eye.  We fitted the source SEDs using the SED models developed by \cite{2006ApJS..167..256R} to obtain estimates of the quantities $M_\star$, the mass of the central star, $\dot M$ the mass accretion/infall rate, $M_{\rm env}$, the mass of the protostellar, infalling envelope, and $M_{\rm disk}$, the mass of any circumstellar disk. We then assign classifications using the following criteria: 
 \begin{itemize}
\item P: protostar (Class 0/I), $\dot M/M_\star \ge 10^{-6}$ and $M_{\rm env}/M_\star \ge 0.05$
\item D: disk  (Class II),  $\dot M/M_\star < 10^{-6}$ and $M_{\rm disk}/M_\star \ge 10^{-6}$
\item S: star (Class III)  $\dot M/M_\star < 10^{-6}$ and $M_{\rm disk}/M_\star <10^{-6}$
\end{itemize}

\noindent
In this manner we determined that 42 sources in the merged list of Table A.1 were protostars. We note that the difference between this estimate and that of \citeauthor{2014ApJ...786...37B} is primarily due to our assignment of a disk classification for most of the flat spectrum sources in the \citeauthor{2014ApJ...786...37B} list.
We further note that sources 1, 2, \& 3 are outside \textit{Herschel} footprint. Their extinctions are derived from Planck. 
One of these we classify as a protostar but it was not included in any of our subsequent analysis. 

\onecolumn
\begin{longtable}{lllllllll}
\caption{Young Stellar Objects in the California Molecular Cloud \label{t:sourcelist}}\\
\hline\hline
{ID} & {RA (J2000)} & {DEC (J2000)} & {$A_{\rm K}\ (\text{error})$} & \multicolumn{2}{c}{\citeauthor{2014ApJ...786...37B}} & \multicolumn{2}{c}{\citeauthor{2013ApJ...764..133H}}  & {Class} \\[.15ex]
\cline{5-6} \cline{7-8} \\[-1.25ex]  
& & & & {ID} & {Class} & {ID} & {Class} & \\
\hline \\[-1.5ex]
\endhead 
\hline
\endfoot
1* & 04 01 24.55 & 41 01 49.00 &  0.819 (0.007) & 1 & I & 0 & -- & P \\
2* & 04 01 34.36 & 41 11 43.00 & 0.934 (0.020) & 2 & II & 0 & -- & D \\
3* & 04 02 29.75 & 40 42 41.90 & 0.313 (0.008) & 139 & II & 0 & -- & D \\
4 & 04 09 02.00 & 40 19 13.10 & 1.190 (0.029) & 140 & I & 1 & I & P \\
5 & 04 09 54.71 & 40 06 39.89 & 4.096 (0.054) & -- & -- & 2 & I & P \\
6 & 04 10 00.64 & 40 02 36.10 & 2.857 (0.054) & 3 & II & -- & -- & D \\
7 & 04 10 02.63 & 40 02 48.20 & 3.051 (0.050) & 4 & I & 3 & 0 & P \\
8 & 04 10 03.43 & 39 04 49.50 & 0.338 (0.004) & 141 & III & -- & -- & S \\
9 & 04 10 04.53 & 40 02 37.50 & 2.415 (0.053) & -- & -- & 4 & 0 & P \\
10 & 04 10 05.62 & 40 02 38.60 & 3.146 (0.062) & 5 & II & 5 & 0 & D \\
11 & 04 10 07.08 & 40 02 34.58 & 2.717 (0.068) & -- & -- & 6 & 0 & P \\
12 & 04 10 08.41 & 40 02 24.40 & 2.621 (0.076) & 6 & I & 7 & I & P \\
13 & 04 10 11.16 & 40 01 26.20 & 2.390 (0.071) & 7 & I & 8 & 0 & P \\
14 & 04 10 24.41 & 38 05 22.70 & 0.435 (0.008) & 142 & II & -- & -- & D \\
15 & 04 10 40.51 & 38 05 00.40 & 1.339 (0.034) & 8 & II & -- & -- & D \\
16 & 04 10 41.09 & 38 07 54.50 & 1.917 (0.090) & 10 & I & 9 & I & P \\
17 & 04 10 41.63 & 38 08 05.80 & 3.277 (0.118) & 9 & II & -- & -- & D \\
18 & 04 10 42.11 & 38 05 59.90 & 1.682 (0.017) & 11 & III & -- & -- & S \\
19 & 04 10 47.61 & 38 03 33.80 & 0.383 (0.010) & 12 & II & -- & -- & D \\
20 & 04 10 49.16 & 38 04 45.80 & 2.714 (0.054) & 13 & II & 10 & F & D \\
21 & 04 12 08.47 & 38 01 46.60 & 0.341 (0.005) & 143 & III & -- & -- & S \\
22 & 04 12 40.54 & 38 14 26.81 & 0.549 (0.007) & -- & -- & 11 & I/0 & P \\
23 & 04 12 57.64 & 39 14 18.30 & 0.293 (0.005) & 144 & III & -- & -- & S \\
24 & 04 13 44.57 & 39 04 35.70 & 0.414 (0.005) & 145 & III & -- & -- & S \\
25 & 04 15 11.20 & 38 39 57.10 & 0.341 (0.005) & 146 & II & -- & -- & D \\
26 & 04 15 54.05 & 38 34 13.10 & 0.396 (0.005) & 147 & III & -- & -- & S \\
27 & 04 17 05.93 & 37 22 18.70 & 0.208 (0.005) & 148 & III & -- & -- & S \\
28 & 04 19 44.67 & 38 11 21.90 & 0.501 (0.006) & 14 & F & -- & -- & D \\
29 & 04 20 52.46 & 38 06 35.80 & 0.604 (0.007) & 15 & III & -- & -- & S \\
30 & 04 21 37.95 & 37 34 41.80 & 3.497 (0.062) & 16 & II & 12 & F & D \\
31 & 04 21 38.08 & 37 35 40.90 & 3.167 (0.050) & 17 & III & -- & -- & S \\
32 & 04 21 40.80 & 37 33 59.00 & 4.874 (0.080) & 18 & I & 13 & I & P \\
33 & 04 23 05.46 & 38 07 36.90 & 0.271 (0.005) & 149 & III & -- & -- & S \\
34 & 04 24 49.34 & 37 16 46.40 & 1.128 (0.008) & 19 & III & -- & -- & S \\
35 & 04 24 59.04 & 37 17 52.91 & 1.307 (0.020) & -- & -- & 14 & I & P \\
36 & 04 25 07.83 & 37 15 19.30 & 2.052 (0.053) & -- & -- & 15 & 0 & P \\
37 & 04 25 38.48 & 37 07 01.20 & 6.295 (0.068) & 20 & I & 16 & I/0 & P \\
38 & 04 25 39.79 & 37 07 08.20 & 6.452 (0.051) & 21 & F & 17 & II & D \\
39 & 04 27 13.74 & 36 27 10.70 & 0.428 (0.003) & 150 & II & -- & -- & D \\
40 & 04 27 50.80 & 36 31 26.40 & 0.523 (0.003) & 22 & II & -- & -- & D \\
41 & 04 27 58.26 & 36 33 26.50 & 0.546 (0.003) & 23 & II & -- & -- & D \\
42 & 04 28 02.89 & 36 40 58.60 & 0.802 (0.004) & 24 & II & -- & -- & D \\
43 & 04 28 15.15 & 36 30 28.60 & 1.262 (0.004) & 25 & F & 18 & F & D \\
44 & 04 28 21.16 & 36 24 47.80 & 0.753 (0.003) & 26 & II & -- & -- & D \\
45 & 04 28 21.36 & 36 30 21.50 & 0.919 (0.003) & 27 & II & -- & -- & D \\
46 & 04 28 35.09 & 36 25 06.40 & 1.961 (0.020) & 28 & I & 19 & I & P \\
47 & 04 28 37.89 & 36 24 55.30 & 1.757 (0.017) & 29 & II & 20 & F & D \\
48 & 04 28 38.56 & 36 25 28.90 & 2.815 (0.021) & 30 & I & 21 & I & P \\
49 & 04 28 43.35 & 36 25 11.70 & 0.887 (0.005) & 31 & II & -- & -- & D \\
50 & 04 28 43.67 & 36 28 39.30 & 1.801 (0.023) & 32 & I & 22 & I/0 & P \\
51 & 04 28 44.43 & 36 24 45.60 & 0.758 (0.004) & 33 & F & -- & -- & D \\
52 & 04 28 49.58 & 36 29 10.70 & 1.165 (0.004) & 34 & I & -- & -- & D \\
53 & 04 28 55.30 & 36 31 22.50 & 1.965 (0.035) & 35 & I & 23 & I & P \\
54 & 04 28 55.56 & 35 24 46.00 & 0.305 (0.004) & 151 & II & -- & -- & D \\
55 & 04 28 59.11 & 36 23 11.20 & 0.442 (0.003) & 36 & II & -- & -- & D \\
56 & 04 29 11.53 & 35 04 49.50 & 0.233 (0.005) & 152 & II & -- & -- & D \\
57 & 04 29 14.38 & 35 15 24.50 & 0.363 (0.004) & 153 & II & -- & -- & D \\
58 & 04 29 39.01 & 35 16 10.50 & 0.459 (0.003) & 37 & II & -- & -- & D \\
59 & 04 29 40.01 & 35 21 08.90 & 0.696 (0.006) & 38 & I & -- & -- & P \\
60 & 04 29 43.58 & 35 13 38.60 & 0.646 (0.004) & 39 & II & -- & -- & D \\
61 & 04 29 44.21 & 35 12 30.00 & 0.612 (0.005) & 40 & F & -- & -- & D \\
62 & 04 29 46.28 & 36 19 23.50 & 0.423 (0.003) & 154 & F & -- & -- & D \\
63 & 04 29 47.28 & 35 10 19.20 & 0.406 (0.003) & 41 & II & -- & -- & D \\
64 & 04 29 47.42 & 35 11 33.50 & 0.605 (0.006) & 42 & II & -- & -- & D \\
65 & 04 29 48.54 & 35 12 12.50 & 0.671 (0.004) & 43 & II & -- & -- & D \\
66 & 04 29 49.21 & 35 14 22.70 & 1.553 (0.019) & 44 & F & -- & -- & D \\
67 & 04 29 49.61 & 35 14 43.80 & 2.186 (0.025) & 45 & II & -- & -- & D \\
68 & 04 29 50.17 & 35 14 44.50 & 2.186 (0.025) & 136 & II & -- & -- & D \\
69 & 04 29 50.84 & 35 15 57.90 & 2.249 (0.013) & 46 & F & -- & -- & D \\
70 & 04 29 51.01 & 35 15 47.50 & 2.249 (0.013) & 47 & I & -- & -- & P \\
71 & 04 29 52.54 & 35 22 23.60 & 0.292 (0.003) & 155 & III & -- & -- & S \\
72 & 04 29 53.46 & 35 15 48.50 & 2.108 (0.013) & 48 & F & -- & -- & D \\
73 & 04 29 54.15 & 35 10 21.60 & 0.457 (0.006) & 49 & F & -- & -- & D \\
74 & 04 29 54.18 & 36 11 57.30 & 0.985 (0.006) & 156 & F & 24 & II & D \\
75 & 04 29 54.79 & 35 18 02.50 & 0.780 (0.006) & 50 & II & 25 & F & D \\
76 & 04 29 56.27 & 35 17 42.90 & 0.760 (0.009) & 51 & I & -- & -- & P \\
77 & 04 29 59.19 & 36 10 16.10 & 0.773 (0.006) & 157 & II & 26 & II & D \\
78 & 04 29 59.76 & 35 13 34.20 & 0.788 (0.006) & 52 & II & -- & -- & D \\
79 & 04 30 00.16 & 36 03 22.70 & 0.531 (0.004) & 53 & II & -- & -- & D \\
80 & 04 30 01.14 & 35 17 24.60 & 0.333 (0.002) & 54 & III & -- & -- & S \\
81 & 04 30 01.52 & 36 07 33.30 & 0.449 (0.003) & 158 & II & -- & -- & D \\
82 & 04 30 01.88 & 35 38 14.70 & 0.350 (0.005) & 159 & II & -- & -- & D \\
83 & 04 30 02.63 & 35 15 14.30 & 0.831 (0.006) & 55 & II & -- & -- & D \\
84 & 04 30 03.63 & 35 14 20.10 & 2.331 (0.016) & 56 & I & -- & -- & P \\
85 & 04 30 04.23 & 35 09 45.90 & 0.523 (0.005) & 57 & II & -- & -- & D \\
86 & 04 30 04.25 & 35 22 23.80 & 0.336 (0.002) & 58 & II & -- & -- & D \\
87 & 04 30 07.43 & 35 14 57.90 & 0.659 (0.005) & 59 & II & -- & -- & D \\
88 & 04 30 07.73 & 35 15 48.40 & 0.429 (0.002) & 60 & II & -- & -- & D \\
89 & 04 30 08.25 & 35 14 10.00 & 1.528 (0.009) & 61 & I & -- & -- & P \\
90 & 04 30 08.74 & 35 14 37.50 & 1.434 (0.010) & 62 & II & -- & -- & D \\
91 & 04 30 09.51 & 35 14 40.30 & 1.101 (0.006) & 63 & I & -- & -- & P \\
92 & 04 30 09.80 & 35 40 35.50 & 0.447 (0.005) & 64 & II & -- & -- & D \\
93 & 04 30 09.80 & 36 13 35.40 & 0.444 (0.003) & 160 & II & -- & -- & D \\
94 & 04 30 09.86 & 35 14 16.30 & 1.528 (0.010) & 137 & II & -- & -- & D \\
95 & 04 30 09.91 & 35 15 53.90 & 0.420 (0.003) & 65 & II & -- & -- & D \\
96 & 04 30 12.34 & 35 09 34.60 & 1.531 (0.015) & 66 & II & -- & -- & D \\
97 & 04 30 13.09 & 35 13 58.60 & 1.165 (0.006) & 67 & II & -- & -- & D \\
98 & 04 30 14.53 & 35 13 32.60 & 1.315 (0.008) & 68 & II & -- & -- & D \\
99 & 04 30 14.74 & 35 20 14.30 & 0.583 (0.004) & 69 & II & -- & -- & D \\
100 & 04 30 14.95 & 36 00 08.50 & 2.047 (0.028) & 70 & I & 27 & I & D \\
101 & 04 30 15.21 & 35 16 39.80 & 1.067 (0.012) & 138 & F & -- & -- & D \\
102 & 04 30 15.76 & 35 56 57.80 & 0.753 (0.020) & 71 & I & 28 & II & D \\
103 & 04 30 16.27 & 35 42 42.90 & 0.424 (0.005) & 72 & II & -- & -- & D \\
104 & 04 30 17.84 & 36 03 26.60 & 0.602 (0.005) & 73 & III & -- & -- & S \\
105 & 04 30 18.08 & 35 45 38.90 & 0.546 (0.008) & 74 & II & -- & -- & D \\
106 & 04 30 18.99 & 35 42 12.00 & 0.479 (0.006) & 75 & II & -- & -- & D \\
107 & 04 30 19.59 & 35 08 21.60 & 1.220 (0.015) & 76 & F & -- & -- & D \\
108 & 04 30 22.19 & 36 04 35.90 & 0.857 (0.008) & 77 & II & -- & -- & D \\
109 & 04 30 22.68 & 35 19 08.10 & 0.617 (0.007) & 78 & II & -- & -- & D \\
110 & 04 30 23.82 & 35 21 12.30 & 0.740 (0.007) & 79 & I & -- & -- & P \\
111 & 04 30 24.33 & 34 59 16.50 & 0.440 (0.006) & 80 & III & -- & -- & S \\
112 & 04 30 24.68 & 35 45 20.60 & 1.221 (0.028) & 81 & I & 29 & I & P \\
113 & 04 30 25.03 & 35 43 17.90 & 1.115 (0.012) & 82 & II & -- & -- & D \\
114 & 04 30 25.89 & 35 48 11.30 & 1.507 (0.019) & 83 & II & -- & -- & D \\
115 & 04 30 27.02 & 35 20 28.40 & 0.811 (0.008) & 84 & II & -- & -- & D \\
116 & 04 30 27.04 & 35 45 50.50 & 2.062 (0.023) & 85 & F & 30 & I & P \\
117 & 04 30 27.41 & 35 09 17.80 & 1.919 (0.077) & 86 & I & 31 & I & P \\
118 & 04 30 27.75 & 35 46 15.00 & 2.629 (0.022) & 87 & F & 32 & II & D \\
119 & 04 30 28.09 & 35 09 16.40 & 2.970 (0.101) & 88 & I & -- & -- & P \\
120 & 04 30 28.42 & 35 32 41.90 & 0.428 (0.005) & 89 & II & -- & -- & D \\
121 & 04 30 28.44 & 35 49 17.60 & 2.590 (0.015) & 90 & F & -- & -- & D \\
122 & 04 30 28.61 & 35 47 40.70 & 1.798 (0.015) & 91 & II & 33 & II & D \\
123 & 04 30 28.71 & 35 47 49.80 & 1.826 (0.018) & 92 & II & -- & -- & D \\
124 & 04 30 28.98 & 35 07 54.00 & 0.769 (0.009) & 93 & II & -- & -- & D \\
125 & 04 30 29.61 & 35 27 17.20 & 0.965 (0.010) & 94 & II & -- & -- & D \\
126 & 04 30 29.66 & 35 06 39.00 & 1.102 (0.010) & 95 & II & -- & -- & D \\
127 & 04 30 30.14 & 35 06 39.20 & 1.154 (0.011) & 96 & II & 34 & II & D \\
128 & 04 30 30.28 & 35 21 04.00 & 0.581 (0.006) & 97 & II & -- & -- & D \\
129 & 04 30 30.43 & 35 18 33.70 & 0.335 (0.003) & 98 & II & -- & -- & D \\
130 & 04 30 30.51 & 35 17 44.70 & 0.383 (0.004) & 99 & II & -- & -- & D \\
131 & 04 30 30.56 & 35 51 44.00 & 3.344 (0.053) & 100 & I & 35 & I & P \\
132 & 04 30 31.58 & 35 45 13.70 & 3.073 (0.039) & 101 & F & 36 & II & D \\
133 & 04 30 32.35 & 35 36 13.40 & 1.062 (0.016) & 102 & II & 37 & F & P \\
134 & 04 30 36.80 & 35 54 36.20 & 4.275 (0.132) & 103 & I & 38 & I & P \\
135 & 04 30 37.40 & 36 00 18.00 & 0.711 (0.009) & 104 & II & -- & -- & D \\
136 & 04 30 37.51 & 35 13 48.60 & 0.460 (0.005) & 105 & II & -- & -- & D \\
137 & 04 30 37.51 & 35 50 31.70 & 4.321 (0.101) & 106 & II & 39 & F & D \\
138 & 04 30 37.89 & 35 51 01.40 & 3.383 (0.080) & 107 & I & 40 & 0 & P \\
139 & 04 30 38.26 & 35 49 59.30 & 2.342 (0.077) & 108 & II & 41 & 0 & P \\
140 & 04 30 38.38 & 35 50 22.60 & 3.720 (0.092) & -- & -- & 42 & 0 & P \\
141 & 04 30 38.65 & 35 54 39.10 & 4.404 (0.162) & 109 & F & 43 & F & D \\
142 & 04 30 39.12 & 35 44 49.80 & 0.726 (0.008) & 110 & II & -- & -- & D \\
143 & 04 30 39.16 & 35 52 03.80 & 2.276 (0.040) & 111 & F & 44 & F & P \\
144 & 04 30 39.31 & 35 52 00.70 & 2.647 (0.037) & 112 & F & -- & -- & D \\
145 & 04 30 39.56 & 35 18 06.90 & 0.337 (0.004) & 113 & II & -- & -- & D \\
146 & 04 30 39.58 & 35 11 12.80 & 0.403 (0.004) & 114 & II & -- & -- & D \\
147 & 04 30 40.05 & 35 42 10.30 & 0.363 (0.006) & 115 & III & -- & -- & S \\
148 & 04 30 40.14 & 35 31 34.10 & 1.497 (0.032) & 116 & II & -- & -- & D \\
149 & 04 30 41.16 & 35 29 41.00 & 4.681 (0.069) & 117 & I & 45 & I & P \\
150 & 04 30 44.23 & 35 59 51.10 & 2.782 (0.061) & 118 & I & 46 & F & D \\
151 & 04 30 44.69 & 35 10 52.10 & 0.370 (0.005) & 119 & II & -- & -- & D \\
152 & 04 30 45.58 & 34 58 08.00 & 0.997 (0.017) & 120 & II & -- & -- & D \\
153 & 04 30 46.25 & 34 58 56.20 & 2.743 (0.096) & 121 & I & 47 & 0 & P \\
154 & 04 30 47.23 & 35 07 43.20 & 0.745 (0.008) & 122 & II & 48 & II & D \\
155 & 04 30 47.57 & 34 58 24.20 & 3.860 (0.143) & 123 & II & -- & -- & D \\
156 & 04 30 47.90 & 34 58 37.31 & 3.860 (0.143) & -- & -- & 49 & 0 & P \\
157 & 04 30 48.52 & 35 37 53.70 & 2.480 (0.051) & 124 & I & 50 & I & P \\
158 & 04 30 48.61 & 34 58 53.50 & 5.167 (0.168) & 125 & I & 51 & F & S \\
159 & 04 30 49.22 & 34 56 10.30 & 1.025 (0.012) & 126 & I & 52 & F & D \\
160 & 04 30 49.33 & 34 50 46.00 & 0.585 (0.009) & 161 & II & -- & -- & D \\
161 & 04 30 49.34 & 35 36 41.90 & 0.808 (0.011) & 127 & II & -- & -- & D \\
162 & 04 30 49.48 & 34 50 56.20 & 0.559 (0.009) & 162 & II & -- & -- & D \\
163 & 04 30 49.68 & 34 57 27.70 & 0.860 (0.021) & 128 & II & 53 & II & D \\
164 & 04 30 50.57 & 35 33 23.50 & 0.864 (0.007) & 129 & II & -- & -- & D \\
165 & 04 30 50.98 & 35 35 54.80 & 0.516 (0.007) & 130 & II & -- & -- & D \\
166 & 04 30 52.08 & 34 50 08.90 & 0.653 (0.010) & 163 & F & 54 & F & D \\
167 & 04 30 53.50 & 34 56 27.40 & 2.593 (0.041) & 131 & I & 55 & I & D \\
168 & 04 30 53.90 & 35 30 11.00 & 1.097 (0.017) & 132 & II & -- & -- & D \\
169 & 04 30 55.01 & 35 30 56.20 & 0.583 (0.006) & 133 & II & -- & -- & D \\
170 & 04 30 55.99 & 34 56 47.80 & 2.140 (0.028) & 134 & I & 56 & I & D \\
171 & 04 30 56.61 & 35 30 04.50 & 1.513 (0.029) & 135 & I & 57 & I & P \\
172 & 04 30 57.19 & 34 53 53.59 & 2.799 (0.058) & -- & -- & 58 & I & P \\
173 & 04 31 14.67 & 35 56 50.60 & 0.633 (0.008) & -- & -- & 59 & I & P \\
174 & 04 32 05.77 & 36 06 37.50 & 0.359 (0.005) & 164 & III & -- & -- & S \\
175 & 04 32 54.31 & 36 04 44.00 & 0.362 (0.005) & 165 & II & -- & -- & D \\
176 & 04 33 03.15 & 36 02 04.50 & 0.360 (0.006) & 166 & II & -- & -- & D \\
177 & 04 34 53.15 & 36 23 27.89 & 2.284 (0.041) & -- & -- & 60 & II & D 
\end{longtable}
\end{appendix}


\end{document}